\documentclass{article}
\usepackage{amssymb,amsmath, amsthm, amsfonts,dsfont}

\usepackage{latexsym,mathtools}
\usepackage{graphicx}

\usepackage{apptools}

\usepackage{color}

\usepackage[all,cmtip]{xy}

\theoremstyle{plain}
\newtheorem{theorem}{Theorem}[section]
\newtheorem{proposition}[theorem]{Proposition}
\newtheorem{lemma}[theorem]{Lemma}
\newtheorem{corollary}[theorem]{Corollary}

\newtheorem{definition}[theorem]{Definition}

\newtheorem{example}[theorem]{Example}

\newcommand{\const}[1]{\overline{#1}}

\let\leq=\leqslant
\let\geq=\geqslant
\let\Box=\square                            

\let\alg=\bm                                    
\let\class=\mathbb                             
\let\mod=\mathfrak							

\let\alg=\mathbf
\let\log=\mathsf

\newcommand\trans[1]{%
	\savestack{\tmpbox}{\stretchto{%
			\scaleto{%
				\scalerel*[\widthof{\ensuremath{#1}}]{\kern-.6pt\bigwedge\kern-.6pt}%
				{\rule[-\textheight/2]{1ex}{\textheight}}
			}{\textheight}%
		}{0.5ex}}%
	\stackon[1pt]{#1}{\tmpbox}%
}

\let\epsilon=\varepsilon
\let\Lambda\varLambda
\let\Gamma\varGamma
\let\Delta\varDelta
\let\Lambda\varLambda
\let\Omega\varOmega
\let\Theta\varTheta
\let\Xi\varXi
\let\Pi\varPi
\let\Sigma\varSigma

\newcommand{\Al}[1][A]{\ensuremath{\mathbf{#1}} }
\newcommand{\Logic}[1][A]{\boldsymbol{\Lambda(}\mathbf{#1}\boldsymbol{)}}
\newcommand{\Hom}{\mathrm{Hom}}                    
\newcommand{\Fm}{\mathbf{Fm}}
\newcommand{\MFm}{\mathbf{MFm}}

\newcommand{\A}{\mathsf{A}}
\newcommand{\E}{\mathsf{E}}

\begin{document}
\title{Axiomatizing logics of fuzzy preferences using graded modalities\thanks{This is a revised and properly extended version of the conference papers \cite{ViEsGo17a} and \cite{ViEsGo18}.}} 
\author{Amanda Vidal$^1$, Francesc Esteva$^2$ and Lluis Godo$^2$ \\ \\
\small $^1$ Institute of Computer Science of the Czech Academy of Sciences, \\ \small Prague, Czech Republic\\
\small {\tt amanda@cs.cas.cz} \\
\small $^2$ Artificial Intelligence Research Institute (IIIA) - CSIC, Barcelona, Spain \\
\small {\tt \{esteva,godo\}@iiia.csic.es} \\
}
\date{}
\maketitle

\begin{abstract} 
The aim of this paper is to propose a many-valued modal framework to formalize reasoning with both graded preferences and propositions, in the style of van Benthem et al.'s classical modal logics for preferences. To do so, we start from Bou et al.'s minimal modal logic over a finite and linearly ordered residuated lattice. We then define appropriate extensions on a multi-modal language with graded modalities, both for weak and strict preferences, and with truth-constants. Actually, the presence of truth-constants in the language allows us to show that the modal operators $\Box$ and $\Diamond$ of the minimal modal logic are inter-definable. 
Finally, we propose an axiomatic system for this logic in an extended language (where the preference modal operators are definable), and prove completeness with respect to the intended graded preference semantics.
\end{abstract}

\section{Introduction}


Reasoning about preferences is a topic that has received a lot of attention in Artificial Intelligence since many years, see for instance \cite{Stanford,DoHuKaPr11,Ka11k}. Two main approaches to representing and handling preferences have been developed:  the relational and the logic-based approaches.

In the classical setting, a (weak) preference binary relation $P \subseteq W\times W$ on a set of alternatives or worlds $W$ is usually modeled as a preorder, i.e.\ a reflexive and transitive relation, 
where $(a,b) \in P$ is understood as {\it $b$ is at least as preferred as $a$}. 

{When preference becomes a matter of degree,}  preference relations can be attached degrees (usually belonging to the unit interval $[0, 1]$) of fulfillment or strength, so they become {\em fuzzy relations}. A weak fuzzy preference relation $P$  on a set  $W$ will be now a fuzzy preorder $P: W \times W \to [0, 1]$, where $P(a, b)$ is interpreted as the degree in which $b$ is at least as preferred as $a$. Given a t-norm $\ast$, a fuzzy relation $P$ is a $\ast$-preorder if it satisfies 
\begin{itemize}
\item reflexivity: $P(a, a) = 1$ for each $a \in W$,  and 
\item $\ast$-transitivity: $P(a, b) \ast P(b, c) \leq P(a, c)$ for each $a, b, c \in W$. 
\end{itemize}

The most influential reference is the book by Fodor and Roubens \cite{FodRou94}, that was followed by many other works like, for example  \cite{DiBaMo07,DiBaMo10,DiMoBa04,DiBaMo08,DiGaMo08}.  In this setting, many questions have been discussed, like e.g.\ the definition of the strict fuzzy order associated to a fuzzy preorder (see for example \cite{Bod081,Bod082,BodDem08,EsGoVi18}).

The basic assumption in logical-based approaches is that preferences have structural properties that can be suitably described in a formalized language. This is the main goal of the so-called \textit{preference logics}, see e.g. \cite{Stanford}. The first logical systems to reason about preferences go back to S. Halld\'en \cite{So57} and to von Wright \cite{vonWright,vonWright72,Liu10}. Other related works are \cite{EnPa06,BeOtRo05}. 
More recently van Benthem et al. in \cite{vBGiRo09} have presented a modal logic-based formalization of representing and reasoning with preferences.
In that paper the authors first define a basic modal logic with two unary modal operators  $\Diamond$ and $\Diamond ^<$, together with the universal and existential modalities, $\A$ and $\mathsf{E}$ respectively, and axiomatize them. Using these primitive modalities, they consider several (definable) binary modalities to capture different notions of preference relations on classical propositions, and show completeness with respect to the intended preference semantics. Finally they discuss their systems in relation to von Wright axioms for \textit{ceteris paribus} preferences \cite{vonWright}. On the other hand, with the motivation of formalizing a comparative notion of likelihood,  Halpern studies in \cite{Hal97} different ways to extend preorders on a set $X$ to preorders on subsets of $X$ and their associated strict orders. He studies their properties and relations among them, and he also provides an axiomatic system for a {\em logic of relative likelihood}, that is proved to be complete with respect to what he calls {\em preferential structures}, i.e. Kripke models with preorders as accessibility relations. All these works relate to the classical (modal) logic and crisp preference (accessibility) relations. 

In the fuzzy ({or graded}) setting,\footnote{In this paper we will be using the term {\em fuzzy} indistinctly to refer to preference relations or propositions valued on the real unit interval $[0, 1]$ or on a finite linearly ordered scale, rather than using the general term {\em many-valued} for the latter case. Indeed, it is commonly accepted within the community of mathematical fuzzy logic to consider the class of fuzzy logics with an underlying notion of {\em comparative truth}, and this is captured by logics of linearly ordered algebras of truth-values, regardless they are finite or not, see e.g. \cite{BC06}.} as far as the authors are aware, there are not many formal logic-based approaches to reasoning with fuzzy preference relations, see e.g.\  \cite{BaEsFoGo01}. 
More recently, in the  first part of \cite{EsGoVi18} we studied and characterized different forms to define fuzzy relations on the set $\cal{P}({W})$ of subsets of ${W}$, from a fuzzy preorder on ${W}$, in a similar way to the one followed in  \cite{Hal97,vBGiRo09} for  classical preorders, while in the second part we have semantically defined and axiomatized several two-tiered graded modal logics to reason about different notions of preferences on crisp propositions, see also \cite{DBLP:conf/ccia/EstevaGV17}.  
On the other hand, in \cite{ViEsGo17a} we considered a modal framework over a many-valued logic with the aim of generalizing Van Benthem et al.'s modal approach to the case of both fuzzy preference accessibility relations and fuzzy propositions. To do that, we first extended the many-valued modal framework of \cite{BoEsGoRo11} for only a necessity operator $\Box$
by defining an axiomatic system with both necessity and possibility operators $\Box$ and $\Diamond$ over the same class of models. {Unfortunately, in the last part of that paper, there is a mistake in the proof of Theorem 3 (particularly, equation (4)). This left open the question of properly axiomatizing the logic of graded preferences defined there.} 

In this paper we address this problem, extending the work developed in \cite{ViEsGo18}. 
We propose an alternative approach to provide a complete axiomatic system for a logic of fuzzy preferences, studying first the logic with reflexive graded preference relations (as in \cite{ViEsGo18}) an later, extending this system with the corresponding strict (irreflexive) preferences.
 Namely, given  a finite MTL-chain $\bf B$ (i.e. a finite totally ordered residuated lattice) as set of truth values, and given an $B$-valued preference Kripke model $(W, P, e)$, with $P$ a fuzzy preorder valued on $A$, we consider the $a$-cuts $P_b$ of the relation $P$ for every $b \in B$, and for each $b$-cut $P_b$, we consider the corresponding modal operators $\Box_b, \Diamond_b$. These operators are easier to be axiomatized than the original $\Box, \Diamond$, since the relations $P_b$ are not fuzzy any longer, but a nested set of classical (crisp) relations. 
 
 The good news are that, in our rich (multi-modal) logical framework, we can show that the original modal operators $\Box$ and $\Diamond$ are definable, and vice-versa if we expand the logic with Monteiro-Baaz's $\Delta$ operator. Thus, we define and axiomatize a conservative extension of the logic where the original operators can be defined using the new graded operators.

The paper is structured as follows. After this introduction, Section \ref{sec:defs1} deals with basic facts on fuzzy preference relations.  
In Section \ref{sec:modalDefs} we present many-valued modal logics in the more general context (over arbitrary finite bounded commutative integral residuated lattices with constants),  and the intended semantics given by \textit{valued Kripke models}. We close an open problem existing in this setting, namely, whether the operations $\Box$ and $\Diamond$ are interdefinable, proving this is the case, and providing the explicit definition of each operator in terms of the other one. This strongly simplifies the symbolic approach to the logic, since it is only necessary axiomatize one of the modal operators to obtain a logic referring to both. In Section
\ref{sec:defs} we show how to adapt the previous general setting to model graded preference relations: we restrict the evaluations to some arbitrary MTL-chain $\bf B$, introduce auxiliary crisp modalities $\Box_b$ and exhibit a complete axiomatization of a conservative extension\footnote{Namely, while the modal language used is larger, the restriction of the logic to the $\{\Box$, $A\}$ fragment coincides with the original one.} of the preference logic studied in \cite{ViEsGo17a}.
In Section \ref{sec:strictPreference} we study the extension of the previous logic with the strict preference modality $\Box^<$, corresponding to the irreflexive restriction of the preference relation associated to the original $\Box$. We propose a complete axiomatization of a conservative extension of this logic (relying again in the $\Box_b, \Box^<_b$ crisp modalities). In Section \ref{sec:axiomNoCuts} we observe how, by the addition to the logic of the so-called Monteiro-Baaz $\Delta$ operation, we can also provide an axiomatization of the original logic of graded preference models pursued in \cite{ViEsGo17a}, without the necessity of additional modal operations. 
 Lastly, in Section \ref{sec:relations}, we discuss different  possibilities to formalize notions of preferences on fuzzy propositions in preference Kripke models.
We finish with some conclusions and open problems.

\section{Preliminaries on fuzzy preference relations}\label{sec:defs1}

In the classical setting, a (weak) preference relation on a set of alternatives $W$  is usually modeled as  preorder  relation (i.e. a reflexive and transitive relation) $P \subseteq W\times W$ by interpreting $(a,b) \in P$ as denoting {\it $b$ is at least as preferred as $a$}. From $P$ one can define three disjoint relations:
\begin{itemize}
\item the {\it strict preference} $P^< = P \cap P^{d}$,
\item the {\it indifference relation} $P^\approx = P \cap P^{t}$, and
\item the {\it incomparability relation} $P^{<>} = P^{c} \cap P^{d}$.
\end{itemize}
where $P^{d} = \{(a,b) : (b,a) \notin P\}$, $P^{t} = \{(a,b): (b,a) \in P\}$ and $P^{c} = \{(a,b) : (a,b) \notin P\}$.
It is clear that $P^<$ is a strict order (irreflexive, antisymmetric and transitive), $P^\approx$ is an equivalence relation (reflexive, symmetric and transitive) and $P^{<>}$ is irreflexive and symmetric. The triple $(P^<, P^\approx, P^{<>})$ is called a {\it preference structure},\footnote{Although in the literature it is more common the notation $(P,I,J)$ to denote preference structures, with our notation we stress that the structure $(P^<, P^\approx, P^{<>})$ is generated by the weak preference (preorder) relation $P$.}  where the initial weak preference relation can be recovered as $P = P^< \cup P^\approx$.

\begin{example}  \label{examp1}
 Let  be $W = \{ \texttt{bf, bm, cf, cm}\}$ a set of alternatives, where $\texttt b$ stands for {\em beach}, $\texttt c$ for {\em countryside}, $\texttt f$ for {\em fish} and $\texttt m$ for {\em meat}. Assume an agent who prefers fish to meat when going to a beach restaurant but prefers meat to fish when going to a countryside restaurant. Moreover, she prefers having fish in  a beach restaurant to having meat in a countryside restaurant. These preferences are modelled by the following preorder, depicted below by means of a $\{0, 1\}$-valued relation:
$$\begin{array}{l|c|c|c|c|}
 P              & \texttt{bf}    &  \texttt bm & \texttt cf  & \texttt cm \\
  \hline \texttt bf   & 1    & 0 & 0 &  0 \\
  \hline \texttt bm & 1 & 1    & 1 & 1 \\
  \hline \texttt cf   & 1 & 0 & 1   &  1\\
  \hline \texttt cm & 1 & 0 & 0 & 1 \\
  \hline
 \end{array} \\
 $$
where $P(a, b) = 1$ means $(a, b) \in P$, and conversely,   $P(a, b) = 0$ means $(a, b) \not\in P$.  
\end{example}

In the fuzzy setting, preference relations can be attached degrees (usually belonging to the unit interval $[0, 1]$) of fulfillment or strength, so they become {\em fuzzy relations}. In this paper we will assume preference degrees are the domain of a {\em finite} and linearly ordered scale ${\bf B} = (B, \leq, 0, 1)$, with $0$ and $1$ being its bottom and top elements respectively. The restriction to finite linearly ordered universes is due to technical reasons, since the axiomatization of modal logics over infinite algebras is either open or partially solved with drawbacks concerning applicability (namely, the require infinitary inference rules, see eg. \cite{HaTe13}, \cite{ViEsGo16}).
Sometimes we will write also ${\bf B} = (B, \land, \lor, 0, 1)$ to emphasize the lattice operations. 


In this paper, we will assume that a weak $\bf B$-valued preference relation $P$  on a set $W$ will be now a fuzzy $\land$-preorder $P: W \times W \to A$, where $P(a, b)$ is interpreted as the degree in which $v$ is at least as preferred as $u$, that is, satisfying: 
\begin{itemize}
\item reflexivity: $P(u, u) = 1$ for each $u \in W$
\item $\land$-transitivity: $P(u, v) \land P(v, w) \leq P(u, w)$ for each $u, v, w \in W$
\end{itemize}

\begin{example} \label{examp2}
The following is a graded refinement of the classical preference relation in Example  \ref{examp1}. Over the same alternatives as above, now the agent slightly prefers having meat to fish when going to a countryside restaurant, while she strongly prefers having fish to meat when going to a beach restaurant.  Also, she usually prefers going to the beach rather than to the countryside. A possible graded modelling of these preferences can be done with the fuzzy preference relation $P: W \times W \to [0, 1]$ defined as follows:

\begin{center}
$\begin{array}{l|c|c|c|c|}
 P              & \texttt bf    & \texttt bm & \texttt cf  & \texttt cm \\
  \hline \texttt bf   & 1    & 0.5 & 0.5 &  0.5 \\
  \hline \texttt bm & 0.8 & 1    & 0.6 & 0.8 \\
  \hline \texttt cf   & 0.8 & 0.5 & 1   &  0.7\\
  \hline \texttt cm & 0.6 & 0.5 & 0.5 & 1 \\
  \hline
 \end{array}  
 $
 \end{center}
 It is easy to check that this is indeed a fuzzy $\land$-preorder, where $\land$ denotes the minimum t-norm. 
\end{example}

As in the classical case, from $P$ it is easy to define $\bf B$-valued relations corresponding to graded counterparts of the strict and indifference relations associated to $P$:\footnote{Valued-based counterparts of the indifference relation $P^{<>}$ have also been defined and discussed in the literature of fuzzy preference relations (see e.g.\ \cite{FodRou94}), but we do not go into further details since this notion will play no role in the logic formalisms we deal with in this paper. }

\begin{itemize}
\item First, we can define the indifference degree between two states, from the preferential point of view, by $P^{\approx}(u,v) \coloneqq P(u,v) \wedge P(v,u)$, providing the degree to which both $u$ is preferred to $v$ and, vice-versa, $v$ is preferred to $u$. This is a $\land$-similarity relation, i.e. a  reflexive, symmetric and $\land$-transitive $\bf B$-valued relation.

\item This allows for defining a second preference relation $P^<$ corresponding to the strict counterpart of $P$ by, roughly speaking, ``removing'' the indifferent pairs of alternatives or worlds from the relation $P$. This amounts to consider $P^<$ as the least $B$-valued relation $R$ such that
 $P = R \vee P^{\approx}$. Taking the point-wise smallest solution of this equation leads to the following definition: 

 \[P^<(u,v) \coloneqq \begin{cases} P(u,v), & \hbox{ if }P(u,v) > P(v, u), \\ 0, & \hbox{otherwise.} \end{cases}\]
 It can be checked that if $P$ is $\land$-transitive, then so is $P^<$ (see e.g. \cite{EsGoVi18}), and thus it can be considered to be a {\em fuzzy strict order}, in the sense that the following counterpart of anti-symmetry property holds for $P^<$: if $P^<(u,v) > 0$ then $P^<(v, u) = 0$. 
 
\end{itemize}

\begin{example} The following are the indifference and strict preference relations corresponding to the fuzzy preference relation in Example \ref{examp2}. 
\begin{center}
$\begin{array}{l|c|c|c|c|}
 P^\approx             & \texttt bf    & \texttt bm & \texttt cf  & \texttt cm \\
  \hline \texttt bf   & 1    & 0.5 & 0.5 &  0.5 \\
  \hline \texttt bm & 0.5 & 1    & 0.5 & 0.5\\
  \hline\texttt  cf   & 0.5 & 0.5 & 1   &  0.5\\
  \hline \texttt cm & 0.5 & 0.5 & 0.5 & 1 \\
  \hline
 \end{array}  
 $  
 \hspace*{1cm}
 $\begin{array}{l|c|c|c|c|}
 P^{< }             & \texttt bf    & \texttt bm & \texttt cf  & \texttt cm \\
  \hline \texttt bf   & 1    & 0 & 0    &  0 \\
  \hline \texttt bm & 0.8 & 1 & 0.6 & 0.8 \\
  \hline \texttt cf   & 0.8 & 0 & 1    &  0.7\\
  \hline \texttt cm & 0.6 & 0 & 0    & 1 \\
  \hline
 \end{array}  
$
 \end{center} \mbox{}
\end{example}

In the next sections we will define and axiomatize a modal preference logic where the initial preorder $P$ together with its corresponding indifference relation $P^\equiv$ and strict preference $P^<$ can be dealt with. To do so, we need to resort the level-cuts of the preference relations and to observe the following facts:

\begin{itemize}
\item Given the initial fuzzy $\land$-preorder $P$, we can define, for each $b \in B$, its corresponding $b$-cut $P_b = \{(u,v) : P(u,v) \geq b\}$, which is a classical preorder.  
\item Analogously, from the corresponding fuzzy strict order $P^<$, we can also define, for each $b \in B$,  the corresponding $a$-cut $(P^<)_b = \{(u,v): P^<(u,v) \geq b\}  = \{ (u, v) : P(u, v) \geq b, P(u, v) > P(v, u) \} $. In this case, the relations $(P^<)_b$ are classical orders. 
\item For each level-cut relation $P_b$ we can also define the corresponding strict order $(P_b)^<$. By definition it is $(P_b)^< = \{(u,v): (u, v) \in P_b, (v, u) \not\in P_b \} = \{(u,v): P(u, v) \geq b, P(v, u) < b \}.$

\item An equivalent expression for $(P_b)^<$ is $(P^<)_b = \{ (u, v) : \exists a \geq b,  P(u, v) \geq a, P(u, v) < a\}$. 

\item Finally, one can also check that  $(P_b)^<$ is always included in $(P^<)_b $, i.e.\ $(P_b)^< \subseteq (P^<)_b $.
\end{itemize}

In general, $(P_b)^<$ and  $(P^<)_b $ do not coincide, as the following example shows.

\begin{example} Let $\bf B$ be the scale where $B = \{0, b, 1\}$ and $0 < b < 1$. 
Let $P$ be the $\bf B$-valued preorder on the universe $W =\{x,y\}$ defined by $P(x,x) = P(y,y)  = P(y,x) = 1$ and $P(x,y)=b$. Then it is obvious that,
\begin{itemize}
\item $P_b = W \times W$ and thus $(P_b)^< = \emptyset$
\item $P^<$ is defined as, $P(u,v) = 1$ if $u = y$ and $v = x$, and $P(u,v)=0$ otherwise. Then $(P^<)_b = \{(y,x)\}$.
\end{itemize}
Thus $(P^<)_b \subsetneq (P_b)^<$.
\end{example}

As usual, one can recover the fuzzy relations $P$ and $P^<$ from their crisp level-cut relations: 
\[P(u,v) = \max \{b \land P_b(u,v) : b \in B\}, \quad   P^<(u,v) = \max \{b \land (P^<)_b(u,v) : b \in B\}\]
Moreover, even if the relations $(P^<)_b$ and $(P_b)^<$ do not generally coincide, $P^<$ can also be  recovered from the crisp relations $\{(P_b)^< : b \in B\}$.


\begin{proposition} \label{prop:twostrictcuts} 
Let $P$ be an $\bf B$-valued $\land$-preorder on a universe $W$. Then for all $u,v \in W$,
\begin{itemize}
\item[] $P^<(u,v) = \max \{b \land (P^<)_b(u,v) : b \in B\} =  \max \{b \land (P_b)^<(u,v) : b \in B\}$
\end{itemize}
\end{proposition}

\begin{proof}
Observe that:
\begin{itemize}
\item If $P^<(u,v) = 0$ then it is easy to check that $(P^<)_b(u,v) = (P_b)^<(u,v) = 0$ for all $b\in A$.
\item  If $P^<(u,v) \neq 0$, then $P^<(u,v) = P(u,v) >P(v,x)$. Then:

 \begin{itemize}
\item For $a = P(u,v)$, it is obvious that $(P^<)_a(u,v) = 1$ and $(P^<)_b(u,v) = 0$ for all $b>a$.
\item By definition, for $a = P(u,v)$ we have $P_a(u,v) = 1$ and $P_a(v,u) = 0$. Then $(P_a)^<(u,v) = 1$ and it is also obvious that $(P_b)^<(u,v) = 0$ for all $b > a$.
 \end{itemize}
 \end{itemize}
Thus the claim is proved.
\end{proof}

\section{Many-valued modal logics: language and semantics }\label{sec:modalDefs}
A suitable formalism over which we can construct a graded preference framework is that of many-valued modal logics. In particular, we take as starting point the modal logic introduced in \cite{BoEsGoRo11} and further studied in \cite{ViEsGo17a}: finitely-valued (propositional) fuzzy logics enriched with modal-like operations.

Let us begin by defining the formal language of our underlying many-valued  propositional setting.
Let $\alg{B} = (B, \land, \lor, \odot, \to, 0, 1)$ be a {\em finite} (bounded, integral, commutative) residuated lattice \cite{GaJiKoOn07}, and consider its canonical expansion $\Al[B^c]$ by adding a new constant $\overline{a}$ for
every element $b \in B$ (canonical in the sense that the interpretation of $\overline{b}$ in $\Al[B^c]$ is $b$ itself). A negation operation $\neg$ can always be defined as $\neg x = x \to 0$. 

The logic associated with $\Al[B^c]$ will be denoted by $\Logic[B^c]$, and the set $\Fm$ of propositional formulas of its language is defined in the usual way from a set of propositional variables $\cal V$ in the language of residuated lattices (we will use the same symbol to denote connectives and operations), 
including constants $\{ \overline{b}: b \in B\}$. The corresponding logical consequence relation $ \models_{\Al[B^c]}$ is defined as follows: for any set of formulas $\Gamma\cup\{\varphi\} \subseteq \Fm$, 
\begin{itemize}
	\item $ \Gamma \models_{\Al[B^c]} \varphi$  if, and only if,  
	\item[] $\forall h\in\Hom(\Fm,\Al[B^c]),$ if $ h[\Gamma] \subseteq \{ 1 \}$ then $ h(\phi)=1$, 
\end{itemize}
where $\Hom(\Fm,\Al[B^c])$ denotes the set of evaluations (homomorphisms) of formulas on $\Al[B^c]$.

Lifting to the modal level, we can expand the propositional language $\Fm$  by modal operators in different ways. The most general way to do so is consider a pair of unary operators $\Box, \Diamond$, and  build the corresponding set $\bf KFm$ of \textit{modal formulas}, again defined as usual from a set $\cal V$ of propositional variables, residuated lattice operations $\{\land, \lor, \odot, \to\}$, truth constants $\{\overline{b} : b \in B\}$, and  modal operators $\{\Box, \Diamond\}$.

We are now ready to introduce $\alg{B}$-valued Kripke models, a generalization to $\alg{B}$ of classical Kripke models.
\begin{definition} \label{AKmodel}
	\textit{A $\alg{B}$-model}  is a triple $\mod{M}=\langle W, P, e\rangle$ such that
	 \begin{itemize} \item $W$ is a set of worlds, 
	 	\item $P\colon W \times W \to B$ is a $B$-valued binary relation between worlds, and 
	 	\item $e\colon W \times \mathcal{V} \to B$ is a world-wise $\bf B$-evaluation of variables.
	
	The evaluation $e$ is uniquely extended to formulas of $\bf KFm$ by using the operations in $\alg{B}$ for what concerns propositional connectives, and letting 
	\begin{align*}
		e(v,\Box \varphi) = & \bigwedge\limits_{w \in W} \{P(v,w) \rightarrow e(w,\varphi)\}  \\
		 e(v,\Diamond \varphi) =&  \bigvee\limits_{w \in W} \{P(v,w)\odot e(w,\varphi)\}
		\end{align*}
	
\end{itemize}
\end{definition}

We will denote by $\class{K}_{\alg{B}}$ the class of all $\alg{B}$-models. Given an $\alg{B}$-model $\mod{M} \in \class{K}_\alg{B}$ and $\Gamma \cup\{ \varphi\} \subseteq {\bf MFm}$, we write $\Gamma \Vdash_{\mod{M}} \varphi$ whenever for any $v \in W$, 
if $e(v, \gamma) = 1$ for all $\gamma \in \Gamma$, then $e(v, \varphi) = 1$ too. Analogously, for $\class{C} \subseteq \class{K}_{\alg{B}}$,  we write $\Gamma \Vdash_{\class{C}} \varphi$ whenever $\Gamma \Vdash_{\mod{M}} \varphi$ for any $\mod{M} \in \class{C}$.

In \cite{BoEsGoRo11}, the $\Box$- fragment of the previous logic was axiomatized, but it was left as an open question how to axiomatize the logic with both $\Box$ and $\Diamond$ operations. That question was addressed in \cite{ViEsGo17a}, where an axiomatic system was proposed and proved complete. Nevertheless, we propose below a new solution to the problem, that also closes an open question: namely, that of the inter-definability of the modal operators in the above valued setting. While it is well known that in classical modal logic both modal operators are inter definable ($\Box \varphi = \neg \Diamond \neg \varphi$ and $\Diamond \varphi = \neg \Box \neg \varphi$), it was not known if something similar happened in valued cases. In particular, since the negation might fail to be involutive (for instance, it is involutive in \L ukasiewicz logic, but not in other well-known fuzzy logics), the classical interdefinition fails. 

Nevertheless, we can prove different equalities, that will serve us to work with the axiomatic systems presented in \cite{BoEsGoRo11} plus a simple definition of the dual operation.

Given two formulas $\varphi, \psi$, we will write $\varphi \equiv_{\class{K}_\alg{B}} \psi$ if and only if for any $\alg{B}$-model $\mod{M}$ and any $v \in W$ it holds $e(v, \varphi) = e(v, \psi)$.

\begin{lemma}\label{lem:equivalence} Let $\bf B$ be a {\em finite} (bounded, integral, commutative) residuated lattice. Then for any $b \in B$ it holds
\[b = \bigwedge_{a \in B} (b \to a) \to a\]
\end{lemma}
\begin{proof} 
On the one hand, by residuation, $b \leq (b \to a) \to a$ for each $a$, since 
$b \leq (b \to a) \to a \text{ iff } b \odot (b \to a) \leq a,$ which is always true. Thus, $b \leq \bigwedge_{a \in B} (b \to a) \to a$.

On the other hand, $(b \to b) \to b = b$, so $\bigwedge_{a \in B} (b \to a) \to a \leq b.$

\end{proof}

\begin{proposition}[Interdefinability]\label{lem:interdef}
 Let $\bf B$ be a {\em finite} (bounded, integral, commutative) residuated lattice.
Then, the following equalities hold:
\begin{align*}
\Box \varphi \equiv_{\class{K}_\alg{B}}&  \bigwedge_{b \in B} (\Diamond (\varphi \to \overline{b}) \to  \overline{b})\\
\Diamond \varphi \equiv_{\class{K}_\alg{B}} & \bigwedge_{b \in B} (\Box (\varphi \to \overline{b}) \to  \overline{b})
\end{align*}

\end{proposition}

\begin{proof}

Is easy to prove that for any $\alg{B}$, the following equalities hold:
\begin{align*}
& - \Box(\bigwedge_{i \in I} \chi_i)  \equiv_{\class{K}_\alg{B}} \bigwedge_{i \in I} \Box\chi_i, \text{ for } I \text{ being a finite set of indexes}\\
& - \Box(\varphi \rightarrow \const{c})  \equiv_{\class{K}_\alg{B}}  \Diamond \varphi \rightarrow \const{c},  \text{ for any constant }\const{c}
\end{align*}
The first one follows from the definition of the evaluation of $\Box$ as a conjunction. The second one  follows from a general property of any residuated lattice (see eg. \cite{JipTsi02}), that states that for any set $X$ of elements of the universe and any other element $y$ \[\bigwedge_{x \in X}(x \rightarrow y) = \bigvee_{x \in X}x \rightarrow y.\]

Concerning the definition of $\Box$ from $\Diamond$, the previous properties imply that, for any $\alg{B}$, 
\begin{align*}
\bigwedge_{b \in B} (\Diamond (\varphi \to \overline{b}) \to  \overline{b}) \equiv_{\class{K}_\alg{B}}  \bigwedge_{b \in B} \Box ((\varphi \to \const{b}) \to \const{b})
 \equiv_{\class{K}_\alg{B}}  \Box (\bigwedge_{b \in B} ((\varphi \to \const{b}) \to \const{b}))
\end{align*}

From  Lemma \ref{lem:equivalence}, we also know that $ \varphi \equiv_{\class{K}_\alg{B}}  \bigwedge_{b \in B} ((\varphi \to \const{b}) \to \const{b})$, so the two formulas evaluate equally in any world of any model. 
Thus, in particular,  for any $\alg{B}$-model, and any $v \in W$, we can conclude

\begin{align*}
e(v, \bigwedge_{b \in B} (\Diamond (\varphi \to \overline{b}) \to  \overline{b}) ) & = \bigwedge_{w \in W} P(v,w) \rightarrow e(w, \bigwedge_{b \in B} (( \varphi \to \const{b}) \to \const{b})) \\
& =  \bigwedge_{w \in W} P(v,w) \rightarrow e(w, \varphi) \\
& = e(v, \Box \varphi). 
\end{align*}

For what concerns the definability of $\Diamond$ from $\Box$, 
we can use Lemma \ref{lem:equivalence} again to get that
\[\Diamond \varphi \equiv_{\class{K}_\alg{B}}  \bigwedge_{b \in B} ((\Diamond \varphi \to \const{b}) \to \const{b}).\] 
From the second property of $\alg{B}$-models above, we can conclude
\[\Diamond \varphi \equiv_{\class{K}_\alg{B}} \bigwedge_{b \in B} (\Box( \varphi \to \const{b}) \to \const{b}).\]

\end{proof}

After the previous results, it turns out that an axiomatic system addressing both $\Box$ and $\Diamond$ operators with their intended semantics for $\class{K}_{\alg{B}}$ can be easily given by adding to the logic $\boldsymbol{\Lambda}(Fr, \alg{B}^c)$ presented in \cite{BoEsGoRo11} the abbreviation 
\[\Diamond \varphi \coloneqq \bigwedge_{b \in B} (\Box (\varphi \to \overline{b}) \to  \overline{b}).\]
We will denote this axiomatic system by $\mathsf{M}_\alg{B}$. See Appendix A for the details on its definition.

\section{Multi-modal preference logic}\label{sec:defs}

Using the previously defined general modal setting, our objective is formalizing a framework to account for graded preferences in the sense of Section \ref{sec:defs1}. Thus, several particularities arise in respect to the previous general case. To do so, first of all we have to require the accessibility relations $R$ in $\alg{B}$-models be $\wedge$-transitive and reflexive, to capture the transitive and reflexive properties of (weak) preference relations. Also, in order to represent preferences between propositions (as opposed to between alternatives), 
for instance in the style of von Wright's treatment of preferences  \cite{vonWright}, it is necessary to introduce in the language modalities $\A$ and $\mathsf E$ corresponding to the universal relation $W \times W$. This is due to the fact that the most common extensions of preference relations on worlds to preference relations on propositions refer to {\em global} conditions, i.e.\ they express a condition to be satisfied by either at least in one world or in all the worlds of the model, see Section \ref{sec:relations} for more details.

While the restriction to transitive and reflexive models can be dealt with in a systematic way, additional operations to refer to the universal modality $\A$ (and its dual $\mathsf E$, obtained by identifying $\A$ with $\Box$ in Proposition \ref{lem:interdef}) require, for technical reasons, to unfold the modality $\Box$ in a family of cut-modalities $\{\Box_b \colon b \in B\}$.  Moreover, we also need to restrict the kind of propositional algebras of evaluation to linearly ordered ones. Thus, from this point on, we assume \begin{center}
\it{$\alg{B}$ to be a linearly ordered finite (integral, commutative) residuated lattice,}
\end{center} 
or equivalently, to be a {\em finite MTL-chain}.These modifications are due to technical reasons in the completeness proof, resulting from the difficulties posed to axiomatize many-valued modal logics with a crisp accessibility relation (necessary in order to get the desired $\A$ modality) over non-linearly ordered algebras.

Thus, let us define by $\bf MFm$ the set of \textit{multi-modal formulas}, again defined as usual from a set $\cal V$ of propositional variables, (binary) residuated lattice connectives $\{\land, \lor, \odot, \to~\}$, truth constant symbols $\{\overline{a} : b \in B\}$ and the family of unary modalities symbols $\{\Box_b \colon b \in B\}$. 

We are now ready to introduce $\alg{B}$-valued preference Kripke models.
\begin{definition} \label{Amodel}
	\textit{A $\alg{B}$-preference model}  is a triple $\mod{M}=\langle W, P, e\rangle$ such that
	 \begin{itemize} \item $W$ is a set of worlds, 
	 	\item $P\colon W \times W \to B$ is a $B$-valued $\wedge$-pre-order, i.e. a reflexive and $\wedge$-transitive $B$-valued binary relation between worlds, and 
	 	\item $e\colon W \times \mathcal{V} \to B$ is a world-wise $\bf B$-evaluation of variables.
	
	The evaluation $e$ is uniquely extended to formulas of $\bf MFm$ by using the operations in $\alg{B}$ for what concerns propositional connectives, and letting for each $b \in B$, 
\[
		e(v,\Box_ab\varphi) =  \bigwedge\limits_{w: P(v,w) \geq b}  \{e(w,\varphi)\}  \]

\end{itemize}
Sometimes we will also write $v \preceq_b w$ for $P(v, w) \geq b$, or even for $P_b(v, w)$. 
\end{definition}

We will denote by $\class{P}_\alg{B}$ the class of $\alg{B}$-preference models. Given an $\alg{B}$-preference model $\mod{M} \in \class{P}_\alg{B}$ and $\Gamma \cup\{ \varphi\} \subseteq {\bf MFm}$, we write $\Gamma \Vdash_{\mod{M}} \varphi$ whenever for any $v \in W$, 
if $e(v, \gamma) = 1$ for all $\gamma \in \Gamma$, then $e(v, \varphi) = 1$ too. Analogously, we write $\Gamma \Vdash_{\class{P}_\alg{B}} \varphi$ whenever $\Gamma \Vdash_{\mod{M}} \varphi$ for any $\mod{M} \in \class{P}_\alg{B}$.

We will give differentiated symbols to  some particular definable modal operators that enjoy a special meaning in our models. Namely:
\begin{itemize}

\item $\A \varphi := \Box_0 \varphi$ and $\E\varphi := \Diamond_0 \varphi$. \vspace{0.1cm}
	
These operators are in fact universal necessity and possibility modal operators respectively, i.e., 
		\[e(v, \A \varphi) = \bigwedge\limits_{w \in W} \{e(w, \varphi)\}, \quad e(v, \E \varphi) = \bigvee\limits_{w \in W}\{e(w, \varphi)\} .\]

\item $\Diamond_b \varphi := \bigwedge _{a \in B} (\Box_b (\varphi \to \const{a}) \to \const{a})$  \vspace{0.1cm}

Simple computations show that, as expected (from Proposition \ref{lem:interdef}), $$e(v,\Diamond_b \varphi) = \bigvee\limits_{w: P(v,w) \geq b} \{e(w,\varphi)\}.$$

	\item $\Box \varphi :=  \bigwedge_{b \in B} \overline{b} \to \Box_b \varphi $ and $\Diamond \varphi :=  \bigvee_{b \in B} \overline{b} \odot \Diamond_b \varphi$. \vspace{0.1cm}
	
	It is easy to check that the evaluation of these operators in a preference model as defined here, coincides with the usual one for fuzzy Kripke models, i.e.,
	\[e(v, \Box \varphi) = \bigwedge\limits_{w \in W} \{P(v,w) \rightarrow e(w, \varphi)\}, \quad e(v, \Diamond \varphi) = \bigvee\limits_{w \in W}\{P(v,w) \odot e(w, \varphi)\}\]
	
\end{itemize}

Regarding the intuitive meaning of modal formulas of the form $\Diamond \varphi$ and $\Box \varphi$, let us first consider the case $\varphi$ is a crisp formula. Then 
the value of a formula $\Diamond \varphi$ in a world/alternative $v \in W$ of the above preference model 
$$e(v, \Diamond \varphi) = \bigvee\limits_{w \in W, e(w, \varphi) = 1}\{P(v,w)\}$$
is the maximum degree in which some alternative where $\varphi$ holds is preferred to $v$. Similarly, a formula $\Box \varphi$ is evaluated to 
 $$e(v, \Box \varphi) = \bigwedge\limits_{w \in W, e(w, \varphi) = 1}\{P(v,w)\}, $$
the minimum of the degrees in which all alternatives where $\varphi$ is true are preferred to $v$. 
These generalize the semantics of classical preference operators, where $\Box \varphi$ is true in a world if in all preferred alternatives $\varphi$ holds, and $\Diamond \varphi$ is true in a world if there is a preferred alternative where $\varphi$ holds.

In the full general case, where formulas $\varphi$ are also valued on arbitrary values of the algebra $\alg{B}$, the values of modal formulas $\Diamond \varphi$ and $\Box \varphi$ take into account both preference degrees and the values of the formula $\varphi$ at each alternative. In particular the value $P(v,w) \odot e(w, \varphi)$  stands for the truth-evaluation of the conjunctive statement ``$w$ is preferred to $v$ and $\varphi$ is true at $w$'', while $P(v,w) \to e(w, \varphi)$ stands for the truth-evaluation of the implicative statement ``if $w$ is preferred to $v$ then $\varphi$ is true at $w$''.

\begin{example} \label{examp4} Let $B = \{0, 0.1, \ldots, 0.9, 1\}$ and let $\alg{B} = (B, \min, \max, \odot, \to, 0, 1)$ be  the MV-chain over $B$ with the standard {\L}ukasiewicz operations $x \odot y = \max(x+y-1, 0)$ and $x \to y = \min(1-x+y, 1)$. Let us consider a multi-modal language $\bf MFm$ built from the set of propositional variables $V = \{c,b,f,m\}$ (standing for ``countryside", ``beach", ``fish" and ``meat" respectively) and truth-constants $\overline{r}$ for each $r \in B$.
 
The preference relation $P$ from Example \ref{examp2} can be used to define an $\bf B$-preference model $\mod{M} = (W, P, e)$, with $W = \{\texttt{cf,cm,bf,bm}\}$, $P$ defined as in  Example \ref{examp2}, and where the evaluation $e: W \times V \to B$ interprets the propositional variables $\{c,b,f,m\}$ to their intended crisp values, that is,  for $\texttt{x} \in \{\texttt{c,b}\}$ and $\texttt{y} \in \{\texttt{f,m}\}$ we have:  
\[
e(\texttt{xy},c) = 
\begin{cases} 
1, & \text{ if  \texttt{x}} = \texttt{c}, \\ 
0, & \text{otherwise}
\end{cases} 
\quad  \quad
e(\texttt{xy}, b) = \begin{cases} 
1, &  \text{ if \texttt{x}} = \texttt{b}, \\ 
0, & \text{otherwise}
\end{cases}  
\]
and similarly for the variables $f$ and $m$.

To see how the the model $\mod{M}$ can be used to evaluate other (fuzzy) propositions taking advantage of the rich algebraic setting, we can consider the propositions  ``light meal", denoted $l$, and ``heavy meal", denoted $h$,  defined as the following compound formulas: 

\begin{center}
$l := (\overline{0.8} \land f) \vee (\overline{0.2} \land m)$ \\
$h := (\overline{0.7} \land m) \vee (\overline{0.3} \land f)$ \\
\end{center}
leading to the following evaluation of these formulas in the four possible worlds of the model:

\begin{center}
$\begin{array}{l|c|c|c|c|}
 e(\cdot, \cdot)                         & l    & h  \\
  \hline \texttt bf   & 0.8    & 0.3  \\
  \hline \texttt bm & 0.2 & 0.7  \\
  \hline \texttt cf   & 0.8 & 0.3 \\
  \hline \texttt cm & 0.2 & 0.7  \\
  \hline
 \end{array}  
 $
\end{center}

\noindent Further,  one can compute in $\mod{M}$, for instance, the degrees to which the modal formulas $\Diamond l$ and $\Diamond h$ hold true in each of the possible worlds of the model. We exemplify below the computations of these values as $\max-\odot$ compositions (denoted $\circ$) of the graded preference relation $P$ (represented as a matrix) with the evaluations $e(\cdot, l)$ and $e(\cdot, h)$ (represented as vectors): 
\begin{itemize}

\item $e(v, \Diamond l) = \bigvee_{w\in W} P(v, w) \odot e(w, l)$

\begin{center}
$\begin{array}{l|c|c|c|c|}
 P              & \texttt bf    & \texttt bm & \texttt cf  & \texttt cm \\
  \hline \texttt bf   & 1    & 0.5 & 0.5 &  0.5 \\
  \hline \texttt bm & 0.8 & 1    & 0.6 & 0.8 \\
  \hline \texttt cf   & 0.8 & 0.5 & 1   &  0.7\\
  \hline \texttt cm & 0.6 & 0.5 & 0.5 & 1 \\
  \hline
 \end{array}  
 $
 $\quad \circ \quad$
 $\begin{array}{l|c|c|c|c|}
 e     & l    \\
  \hline \texttt bf   & 0.8     \\
  \hline \texttt bm & 0.2   \\
  \hline \texttt cf   & 0.8  \\
  \hline \texttt cm & 0.2   \\
  \hline
 \end{array}  
 $
 $\quad = \quad$
$\begin{array}{l|c|c|c|c|}
 e     & \Diamond l    \\
  \hline \texttt bf   & 0.8     \\
  \hline \texttt bm & 0.6   \\
  \hline \texttt cf   & 0.6  \\
  \hline \texttt cm & 0.4  \\
  \hline
 \end{array}  
 $

 \end{center}

\item $e(v, \Diamond h) = \bigvee_{w\in W} P(v, w) \odot e(w, h)$

\begin{center}
$\begin{array}{l|c|c|c|c|}
 P              & \texttt bf    & \texttt bm & \texttt cf  & \texttt cm \\
  \hline \texttt bf   & 1    & 0.5 & 0.5 &  0.5 \\
  \hline \texttt bm & 0.8 & 1    & 0.6 & 0.8 \\
  \hline \texttt cf   & 0.8 & 0.5 & 1   &  0.7\\
  \hline \texttt cm & 0.6 & 0.5 & 0.5 & 1 \\
  \hline
 \end{array}  
 $
 $\quad \circ \quad$
 $\begin{array}{l|c|c|c|c|}
 e     & h    \\
  \hline \texttt bf   & 0.3     \\
  \hline \texttt bm & 0.7   \\
  \hline \texttt cf   & 0.3  \\
  \hline \texttt cm & 0.7   \\
  \hline
 \end{array}  
 $
 $\quad = \quad$
$\begin{array}{l|c|c|c|c|}
 e     & \Diamond h    \\
  \hline \texttt bf   & 0.3     \\
  \hline \texttt bm & 0.7   \\
  \hline \texttt cf   & 0.4  \\
  \hline \texttt cm & 0.7  \\
  \hline
 \end{array}  
 $
\end{center}
\end{itemize}
\end{example}

\subsection{Axiomatizing fuzzy (weak) preference models}\label{sec:axiomatization}

In this section we axiomatize the logic whose semantics is given by the class $\class{P}_\alg{B}$ of $\alg{B}$-preference models, and based on the use of the graded modalities $\Box_b$ (and in some cases, also the abbreviation $\Diamond_b$), with $b \in B$, introduced above. 
We will denote by $B^+$ the set of positive elements of $\bf B$, namely, $B \setminus \{0\}$.

\begin{definition}\label{def:modalLogic}
We define the fuzzy multi-modal logic $\mathsf{mM}_\alg{B}$ by the following axioms and rules: 
\begin{itemize}
\item Logic $\mathsf{CM}_\alg{B}$ (Appendix A) for each $\Box_b$ with $b \in B$. (This is the axiomatic system of the minimal modal logic over crisp $\alg{B}$-models (\cite{BoEsGoRo11}, see Appendix A for details).

\item For each $a,b \in B$ such that $a \leq b$, nestedness axioms 
\[\Box_a \varphi \rightarrow \Box_b \varphi\]
\item For each $a,b \in B$, reflexivity and transitivity axioms, namely
\[ T_a\colon  \Box_a \varphi \to \varphi, \qquad 4_{a,b} \colon \Box_{a\wedge b} \varphi \to \Box_a \Box_b \varphi\] 
\item Symmetry axiom for $\Box_0$, namely $B_0 \colon \varphi \to \Box_0 \Diamond_0 \varphi$;

\item Modus Ponens rule and the  necessitation rule for each $b\in B$,\footnote{Observe that in $\mathsf{K}_{B}$, due to the inclusion axioms, the necessitation rules for $\Box_b$ for $a \in B^+$ are derivable from the one for $\Box_0$. }  namely 

\begin{center}
$N_{\Box_b} \colon$ from $ \varphi$ derive $ \Box_b \varphi .$
\end{center}

\end{itemize}
It will be also useful later to consider the system $\mathsf{mM}^-_{\alg{B}^+}$ 
obtained from $\mathsf{mM}_\alg{B}$ by dropping the following axioms: \begin{itemize}
\item the reflexivity axioms $T_b$, for $b\in B$,
\item any axiom involving the subindex $0$ (an element not in $B^+)$. 
\end{itemize}

\end{definition}

Note that, for $b > 0$, $\Box_b$ (and so $\Diamond_b$) are graded counterparts of S4 modalities, while $\Box_0$ (and so $\Diamond_0$) is an S5 modality, see e.g. \cite{HC} for a monograph on classical modal logics and their main kinds of modalities.
	
Let $\vdash_{\mathsf{mM}_\alg{B}}$ be the notion of proof for the previous axiomatic system, defined as usual. We can now show that it is indeed complete with respect to our intended semantics given by the class of preference structures $ \Vdash_{\class{P}_\alg{B}} $.  
	\begin{theorem}\label{theorem:completenessM}
		For any $\Gamma, \varphi \subseteq {\bf MFm}$, 
		\begin{enumerate}
			\item[] $\Gamma \vdash_{\mathsf{mM}_\alg{B}} \varphi$ if and only if
			$\Gamma \Vdash_{\class{P}_\alg{B}} \varphi$.
		\end{enumerate}
	\end{theorem}
	\begin{proof}
		Soundness (left to right direction) is easy to check. For what concerns completeness (right to left direction), we can define a canonical model as in \cite{BoEsGoRo11}, 
		$\mathfrak{M}^c = (W^c, \{P^c_b\}_{b \in B}, e^c)$ with a set of crisp accessibility relations as follows, where $Th(\mathsf{mM}_\alg{B}) = \{ \varphi : \;  \vdash_{\mathsf{mM}_\alg{B}} \varphi \}$ denotes the set of theorems of $\mathsf{mM}_\alg{B}$: 
	\begin{itemize}
		\item $ W^c = \{v \in Hom(\MFm, \alg{B}) \colon v(Th(\mathsf{mM}_\alg{B})) =\{1\}\}$,
		\item $P^c_b(v,w)$ if and only if $v(\Box_b \varphi) = 1 \Rightarrow w(\varphi) = 1$ for all $\varphi \in {\bf MFm}$,
		
		\item $e^c(v, p) = v(p)$, for any propositional variable $p$.
	\end{itemize}

To proceed with the completeness proof, it is necessary to prove the so-called Truth Lemma, which states that the evaluation of modal formulas in the model is compatible with the intended semantics. Namely, we have to show that $$e^c(v, \Box_b \varphi) = v(\Box_b \varphi)$$ for any $\varphi$ and any $b \in B$. This is proven in \cite{BoEsGoRo11}, see Appendix A for details.

Next we show that the set $\{P^c_b : b \in B\}$ is a nested set of reflexive and transitive relations. That $P^c_b \subseteq P^c_a$ if $a \leq b$ directly follows from the nestedness axioms, and that each relation $P^c_b$ is reflexive and transitive follows from axioms $T_b$ and $4_{a,b}$. 

Now, from the (crisp) relations $\{P^c_b : b \in A\}$, let us define the fuzzy relation $P^c$ as follows: 
 \[P^c(v, w)  =  \max\{b \in B\colon P^c_b(v, w)\}.\]

It is clear that $P^c(w, v) \geq b$ if and only if $P^c_b(v,w)$. Then, the Truth Lemma for the previous Canonical Model immediately implies 
\[e^c(v, \Box_b \varphi) = \bigwedge\limits_{w \in W^c, P^c(v,w) \geq b}w(\varphi), \]
It follows from axioms $T_b$ that each $P^c_b$ is reflexive, and so, $P^c$ is a reflexive relation as well.
Moreover, from axioms $4_{a,b}$, we get that $P^c$ is $\wedge$-transitive.

The structure $(W^c, P^c, e^c)$ is {\em almost} an $\bf B$-preference model: $P^c_0$ might be a proper subset of $W^c \times W^c$, and not the universal relation. Indeed, observe that, thanks to axioms $T_0, 4_{0,0}$ and $B_0$, $P^c_0$ can be proven to be an equivalence relation, even though it is not necessarily the case that $P^c_0 = W^c \times W^c$. 
Hence, the only remaining step is to show that we can obtain an equivalent model (in the sense of preserving the truth-values of formulas) in which $P^c_0$ is the universal relation, and thus to really get that $\Box_0$ and $\Diamond_0$ are universal modalities. 
 Nevertheless, since $P^c_b \subseteq P^c_0$ for all $b \in B$, for any $v \in W^c$ we can define a restricted model $\mod{M}^c_v = (W^c_v, P^c_v, e^c_v)$  where $W^c_v = \{u \in W^c \colon P^c_0(v,u)\}$, $P^c_v$ is the restriction of $P^c$ to $W^c_v \times W^c_v$, and, for any $u \in W^c_v$ and any formula $\varphi \in {\bf MFm}$, 
\[e^c_v(u, \varphi) = e^c(u, \varphi)  .\]
Now, this model $\mod{M}^c_v$ is indeed a $\bf B$-preference model and thus it belongs to the class $\class{P}_\alg{B}$. 

To conclude the proof, observe that, 
 if $\Gamma \not \vdash_{\mathsf{mM}_\alg{B}} \varphi$, then  there is 
$v \in W^c$ such that $v([\Gamma]) \subseteq\{1\}$ and $v(\varphi) <1 $ (because the modal inference rules affect only theorems of the logic). 
Then, all the previous considerations allow us to prove that, in the model $\mod{M}^c_v$, we have $e^c_v(v, [\Gamma]) \subseteq 1$ and $e^c_v(v, \varphi) < 1$. Hence, $\Gamma \not \Vdash_{\mod{M}^c_v} \varphi$ and thus $\Gamma \not\Vdash_{\class{P}_\alg{B}} \varphi$ as well, and this concludes the completeness proof.

\end{proof}

\section{Adding strict preferences}\label{sec:strictPreference}

As it has been mentioned before, 
in order to provide a framework allowing a finer handling of preference relations, it would be desirable to have a richer language able to also represent strict preference relations between states. 

Within the setting developed in the previous sections, this amounts to consider in the language new modalities and in the models, besides $\bf B$-valued (weak) preference relations on worlds, their strict counterpart. 
Namely, 
given an $\alg{B}$-preference model $\langle W, P, e\rangle$, recall the relation $P^< \colon W \times W \rightarrow B$, the fuzzy strict counterpart of $P$  defined

in Section \ref{sec:defs1}: 

 \[P^<(v,u) \coloneqq 
 \begin{cases} 
 P(v,u) & \hbox{ if }P(v,u) > P(u,v), \\ 
 0 & \hbox{otherwise.} 
 \end{cases}\]
Then, a richer set of formulas, including (fuzzy) modalities for strict preferences $\Box^<, \Diamond^<$, can be evaluated in a $\alg{B}$-preference model $\langle W, P, e\rangle$ relying on the strict preference relation $P^<$, as it was done for $\Box, \Diamond$ formulas over $\alg{B}$-preference models, namely: 

\[e(v, \Box^< \varphi) = \bigwedge_{w \in W} P^<(v,w) \rightarrow e(w, \varphi), \text{ and } e(v, \Diamond^<\varphi) = \bigvee_{w \in W} P^<(v,w) \odot e(w, \varphi)\]

As in Proposition \ref{lem:interdef}, $\Box^<$ and $\Diamond^<$ are inter-definable, so we will mainly work with the $\Box$ modalities, and use the abbreviation
\[\Diamond^<\varphi \coloneqq \bigwedge_{b \in B} (\Box^< (\varphi \to \overline{b}) \to  \overline{b})\]

In the previous section, we relied on the level-cut relations $P_b$, the S4 modalities $\Box_b$ and the universal modality $\Box_0$ to get an indirect axiomatization (the logic $\mathsf{mM}_\alg{B}$) of the graded preference modality $\Box$ and the universal preference $\A$. We follow a similar approach in this section and consider cut strict modalities ${\Box_b^<}$ for $b > 0$. These modalities are to be interpreted by transitive and irreflexive relations.  However, the addition to the system $\mathsf{mM}_\alg{B}$ of these modalities in such a way that the new system keeps being complete with respect to the intended semantics (that is, models where the relations that evaluate the strict modalities are irreflexive counterparts of the relations that evaluate the S4 modalities) is not immediate.
Indeed,  it is well known that an irreflexive modality cannot be axiomatized by a usual axiom or rule schemata (meaning that there is not an axiom or rule closed under arbitrary substitutions that exactly characterizes the irreflexive models) \cite{BlRiVe01}. Thus, more involved techniques have been developed for this purpose  \cite{Se71,Ga81b}. We will resort here to the {\em bulldozing} construction that, in the classical setting, transforms a reflexive and transitive model into a irreflexive and transitive one with an equivalent logical behavior.

We will see in Section \ref{sbs:completeness} and in Appendix \ref{bulldozed} how this classical construction keeps working in the finite-valued case. 

Nevertheless, the full proof of completeness does not directly follow from the one done for the classical case \cite{vBGiRo09}): although the level-cut accessibility relations are crisp, the values of the formulas at each world are many-valued, posing additional problems to solve.

\subsection{Language and semantics}

Let ${\bf PFm}$ be the expanded set of {graded preference formulas}

defined as usual from a set $\cal V$ of propositional variables, residuated lattice operations $\{\land,\lor,\odot,\to~\}$, truth constants $\{\overline{b} : b \in B\}$, plus modal operators $\{\Box_b \colon b \in B\}$ and $\{{\Box_b^<} \colon b \in B^+\}$. 

The interpretation of the $\Box_b$ modalities will be exactly the same as in Section \ref{sec:defs}, that is, given a  an $\alg{B}$-preference model $\mathfrak{M} = \langle W, P, e\rangle$, we let
\[e(v, \Box_b\varphi) = \bigwedge_{w: P(v, w) \geq b} e(w,\varphi), \qquad 
\]

Regarding the new modalities, a first decision that must be taken is choosing the evaluation of the ${\Box_b^<}$ modalities.  

As discussed in Section \ref{sec:defs}, there are two possible ways to approach the definition of the strict relations starting from the original fuzzy relation $P$: either with the $(P_b)^<$'s, the strict versions of the $b$-cuts of $P$,  or with the $(P^<)_b$'s, the $b$-cuts of the strict version of $P$. As it is shown in Prop.\ \ref{prop:twostrictcuts}, the original $P^<$ can be recovered from both families, which allows to define $\Box^<, \Diamond^<$ using either of the two semantics for ${\Box_b^<},\Diamond^<_b$ (see Lemma \ref{lem:recovercrispmod} below). We will present in this section an axiomatization of the logic using the  family of crisp relations $(P_b)^<$ for each $b \in B^+$.

Therefore, given the model $\mathfrak{M} = \langle W, P, e\rangle$, we define
\[e(v, {\Box_b^<}\varphi) = \bigwedge_{w:v\prec_b w}e(w,\varphi)
\]
where, for any $v, w \in W$ and $a \in B^+$,  $v \prec_b w $ stands for $(P_b)^<(v, w)$, that is, $P(v,w) \geq b$ and $P(w,v) < b$. In terms of the notation $\preceq_b$ introduced from Definition \ref{Amodel}, this is equivalent to say that 
$v \prec_b w$ if and only if $v \preceq_b w $ and $w \not \preceq_b v$.

We will keep denoting by $\Vdash_{\class{P}_{\alg{B}}}$ the logical consequence relation over the extended language ${\bf PFm}$, defined exactly  as done for ${\bf MFm}$ in Section \ref{sec:defs}.

As it happened in the previous section, the graded modality $\Box^<$, and the corresponding $\Diamond_b^<$ with the intended meaning can be defined from the new set of operations. Namely, we consider the following abbreviations in our language:
\begin{itemize}
\item $\Diamond^<_b \varphi \coloneqq \bigwedge_{b \in B} ({\Box_b^<} (\varphi \to \overline{b}) \to  \overline{b})$;
\item $\Box^< \varphi \coloneqq \bigwedge_{b \in B} b  \rightarrow {\Box_b^<} \varphi$.
\end{itemize}

It follows from Proposition \ref{lem:interdef} that, under the above definition, in any preference model we get 
$e(v, \Diamond^<_b \varphi) = \bigvee_{w:v\prec_aw}e(w,\varphi)$.

On the other hand, next lemma shows that the definition of $\Box^<$ in the above terms is accordance with the intended meaning stated at the beginning of this section.  

\begin{lemma}\label{lem:recovercrispmod}
For any $\alg{B}$-preference model $\mod{M}$ and $v \in W$, it holds
\[e(v, \Box^< \varphi) = \bigwedge_{w \in W}P^<(v,w) \rightarrow e(w, \varphi).\]

\end{lemma}
\begin{proof}
$\bigwedge_{w \in W}P^<(v,w) \rightarrow e(w, \varphi) = \bigwedge_{w \in W} \bigvee\limits_{b \in B^+}\{b \colon P^<_b(v,w)\} \rightarrow e(w, \varphi)$ from Proposition \ref{prop:twostrictcuts}. This equals $\bigwedge_{w \in W} \bigvee\limits_{b \in B^+} b \cdot (v \prec_b w) \rightarrow e(w, \varphi)$ understanding $\prec_b$ as a $\{0,1\}$-valued relation. By properties of residuated lattices, the previous coincides with $\bigwedge_{w \in W} \bigwedge\limits_{b \in B^+}(b \rightarrow ((v \prec_b w) \rightarrow e(w, \varphi)))$. Since the infima are independent, we can swap them and get the independent element out of the interior one to get 
$\bigwedge\limits_{b \in B^+} (b \rightarrow \bigwedge_{w \in W}((v \prec_b w) \rightarrow e(w, \varphi)))$ which is exactly $\bigwedge\limits_{b \in B^+} (b \rightarrow e(v, {\Box_b^<}\varphi))$.

\end{proof}

\subsection{Axiomatization}

In this section we  axiomatize $\Vdash_{\class{P}_\alg{B}}$ over ${\bf PFm}$ using the systems introduced in Definition 

\ref{def:modalLogic}. Note that, as we commented above, the modalities $\Box^<_0$ and $\Diamond^<_0$ will be omitted, and that is the reason behind removing in the previous definition all axioms concerning the value $0$.

\begin{definition}\label{def:prefLogic}
We define the fuzzy preference logic $\mathsf{P}_\alg{B}$ by the following axioms and rules: 
\begin{itemize}
\item System $\mathsf{mM}_{\alg{B}}$ from Definition \ref{def:modalLogic} for the modalities $\Box_b,$ with $b \in B$;
\item System $\mathsf{mM}^-_{\alg{B}^+}$ from Definition \ref{def:modalLogic} for the modalities  ${\Box_b^<}$ with $b \in B^+ = B \smallsetminus \{0\}$;
\item For each $b \in B^+$, Inclusion axioms $\Box_b \varphi \rightarrow {\Box_b^<}\varphi $;
	
		\item For each $a, b \in B^+$ such that $b \leq a$, Interaction 1 and 2 axioms:  
		\begin{align*}
		(I1) &\qquad  {\Box_b^<} \varphi \rightarrow \Box_a{\Box_b^<}\varphi\\
		(I2) &\qquad {\Box_b^<} \varphi \rightarrow {\Box_b}^{<}\Box_a\varphi
		\end{align*}
		
		\item For each $b \in B^+$ and $a \in B$, Interaction 3 axiom:
		\begin{align*}
		(I3) &\qquad {\Box_b^<}\varphi \wedge (\psi \rightarrow \const{a})  \rightarrow \Box_b (\varphi \vee (\Box_b \psi \rightarrow \const{a})) 
		\end{align*}
	\end{itemize}
\end{definition}

Soundness of $\mathsf{P}_\alg{B}$ with respect to the intended semantics $\Vdash_{\class{P}_\alg{B}}$ is not hard to check. The inclusion axioms follow immediately from the fact that $\prec_b \; \subseteq \; \preceq_b$. 
Let us show soundness of the other axioms.
\begin{lemma}\label{lem:sound12}
Interaction 1 and 2 axioms are valid in $\Vdash_{\class{P}_\alg{B}}$.
\end{lemma}
\begin{proof}

 Assume $b \leq a \in B^+$, and consider any $\alg{B}$-preference model $\mod{M}$, and any $v,w ,u \in W$. 
 Assume $v \preceq_a w$ and $w\preceq_b u$. From $\wedge$-transitivity of $\preceq$ we get that $v\preceq_{a\wedge b} u$, and since $b \leq a$, $v\preceq_b u$
 
If moreover it holds that $w \prec_b u$, by definition it means that $P(u,w) < b \leq a$. Using reflexivity and $\wedge-$transitivity of $P$ it follows that $P(u,v) \wedge P(v,w) \leq P(u,w) < b$. Since $P(v,w) \geq a \geq b$, necessarily $P(u,v) < b$, so by definition, $v \prec_b u$. This proves the Interaction 1 cases. The proof of soundness of Interaction 2 is analogous.
\end{proof}

\begin{lemma}\label{lem:sound3}
	Interaction 3 axiom  is valid in $\Vdash_{\class{P}_\alg{B}}$. 
\end{lemma}
\begin{proof}
Consider a preference model $\mod{M}$, and any $v \in W$. By definition, 
$e(v, \Box_b (\varphi \vee (\Box_b \psi \rightarrow \const{c}))) = \bigwedge\limits_{w:v\preceq_b w} e(w,  (\varphi \vee (\Box_b \psi \rightarrow \const{c})))$
This infimum can be naturally divided in 
\[\bigwedge\limits_{w:v\prec_bw} e(w,  (\varphi \vee (\Box_b \psi \rightarrow \const{c}))) \wedge \bigwedge\limits_{w:v\preceq_bw,w \preceq_b v} e(w,  (\varphi \vee (\Box_b \psi \rightarrow \const{c}))).\]

Concerning the first expression, by monotonicity it is greater or equal than $\bigwedge\limits_{w:v\prec_bw} e(w,  \varphi) = e(v, {\Box_b^<}\varphi)$.

Similarly,  the second expression is greater or equal than 
$\bigwedge\limits_{w:v\preceq_b w, w\preceq_b v} e(w,  \Box_b \psi \rightarrow \const{c})$.
By using the definition and applying some lattice basic results, we get the following chain of (in)equalities:
\begin{align*}
\bigwedge\limits_{w:v\preceq_b w, w\preceq_b v} e(w,  \Box_b \psi \rightarrow \const{c}) &= \bigwedge\limits_{w:v\preceq_b w, w\preceq_b v} ((\bigwedge\limits_{u:w \preceq_b u}  e(u, \psi)) \rightarrow c) \\
& = \bigwedge\limits_{w:v\preceq_b w, w\preceq_b v} (\bigvee\limits_{u:w\preceq_b u}  e(u, \psi) \rightarrow c) \\
&\geq \bigwedge\limits_{w:v\preceq_b w, w\preceq_b v} e(v, \psi) \rightarrow c \\
&= e(v, \psi \rightarrow \overline{c}).
\end{align*}
Then, $e(v, \Box_b (\varphi \vee (\Box_b \psi \rightarrow \const{c}))) \geq e(v, {\Box_b^<}\varphi) \wedge e(v, \psi \rightarrow \overline{c})$, proving the lemma.

\end{proof}

\subsection{Completeness}\label{sbs:completeness}

To prove completeness of $\mathsf{P}_\alg{B}$ with respect to $\Vdash_{\class{P}_\alg{B}}$, we define the canonical model putting together the two sets of modalities in a similar way as it was done for only $\Box_b, \Diamond_b$ in Section \ref{sec:axiomatization}. That is to say, we let $\mod{M}^c = (W^c, \{P_b: b \in B\}, \{P^<_b: >b \in B^+\}, e)$ be the model\footnote{In order to lighten the notation we omit the subscript $c$ in the elements of the canonical model.}  defined by: 
\begin{itemize}
	\item ${W}^c \coloneqq \{h \in Hom({\bf PFm}, \alg{B}) \colon h(Th(\mathsf{P}_\alg{B}))\}$, 
	\item $P^c_b(v,w)$ iff $v(\Box_b \varphi) = 1 \Rightarrow w(\varphi) = 1$, for each $b \in B$, 
	\item ${P}^{c<}_b(v,w) $ iff $ v({\Box_b^<} \varphi) = 1 \Rightarrow w(\varphi) = 1$, for each $b \in B^+$, 
	\item $e^c(v,p) = v(p)$ for all propositional variable $p$.
\end{itemize}
Recall that it can be easily proven that $P^c_b(v,w)$ if and only if $v(\Box_b \varphi) \leq  w(\varphi)$ and $v(\Diamond_b^\leq \varphi) \geq  w(\varphi)$, and the analogous holds for $P^{c<}_b(v,w)$.

It is our objective to prove both, the Truth Lemma (i.e., that $e^c(v,\varphi) = v(\varphi)$ for any formula $\varphi$ in $\bf PFm$), and to see that the evaluation on the previous model coincides with that over the corresponding cut $\alg{B}$-preference model (i.e., only with $P_b$).

The same proof developed in the previous section for the completeness of $\mathsf{mM}_\alg{B}$ (Theorem \ref{theorem:completenessM}) shows the Truth Lemma (for both sets of modalities $\Box_b$ and ${\Box_b^<}$).

Inclusion and nestedness axioms imply that $P^{c<}_b(v,w) \subseteq P^c_b(v,w) \subseteq P^c_0$ for each $b \in B$. Then, as it was done in  $\mathsf{mM}_\alg{B}$, we can restrict $\mod{M}^c$ in such a way that $P^c_0$ is the universal relation, and so get that $\Box_0$ (and  $\Diamond_0$) are universal modalities.

Then, to prove completeness, for a given $v \in \mod{M}^c$, we need to provide an $\alg{B}$-preference model equivalent  to $\mod{M}^c$ at $v$. This amounts to transform the canonical model to an equivalent one, in which the following conditions are equivalent:

\begin{enumerate}
	\item[C1] $ P^{c<}_b (v,w) $ 
	\item[C2] a) $P^c_b(v,w)$,\\
	b) not $P^c_b(w,v)$.
\end{enumerate}

It is easy to see that in the original canonical model, C1 implies C2-a thanks to the inclusion axiom. Let us further see how C2 implies C1. While in the classical approach this can be done by directly relying on Sahlqvist theory (see \cite[Fact 4.]{vBGiRo09}), for the many-valued case such theory has still to be developed and we need to do some calculations.

\begin{lemma}\label{lemma:twoImpliesone}
In the canonical model, C2 implies C1 for any $v,w \in W$.
\end{lemma}
\begin{proof}
	Assume  $P^c_b(v,w)$ and not $P^c_b(w,v)$. Condition C2 implies, by  definition, that there is some formula $\psi$ such that $w(\Box_b \psi) = 1$ and $v(\psi) = \alpha < 1$. Consider now any formula $\varphi$ such that $v({\Box_b^<} \varphi) = 1$, and it is our goal to see that $w(\varphi) = 1$ too. 
	
	Observe that in the previous situation, $v({\Box_b^<} \varphi \wedge (\psi \rightarrow \const{\alpha})) = 1$. Then, by Interaction 3 axiom, we get that $v(\Box_b (\varphi \vee (\Box_b \psi \rightarrow \const{\alpha}))) = 1$ too. Since we assumed that $P_b(v,w)$, this implies that $w(\varphi \vee (\Box_b \psi \rightarrow \const{\alpha})) = 1$. But we know that $w(\Box_b \psi) = 1$, so $w(\Box_b \psi \rightarrow \const{\alpha}) = \alpha < 1$. Since $\alg{B}$ is linearly ordered, this implies that necessarily $w(\varphi) = 1$.	
\end{proof}

To proceed, we need to check that C1 implies C2-b, which is a certain irreflexivity condition. To do so, we will use the bulldozing method to transform the canonical model to an irreflexive one while maintaining its behavior in all other aspects relevant to the proof. The proof is similar to the classical one, but taking into account several accessibility relations at once and the order between them (namely, $P^c_a \subseteq P^c_b$ for all $b \leq a$).

It is worth to point out that in order to be able to proceed with the bulldozing construction, the $\wedge$-transitivity of $P$ in the preference models plays a crucial role, since it is then the case that in our intended models, not only $P_b$ are transitive, but also $P_b^<$. The soundness of this property is necessary in order to successfully unravel the canonical model to an irreflexive one, as we will see below.

It is not only the case that the $P^{c<}_b$'s are $\wedge$-transitive, but also some other properties for them can be proven before proceeding. Namely, observe that  after applying the necessitation rule, axiom $K$ and  Interaction 2 axiom, we get that the formula $\Box_b {\Box_b^<} \varphi \rightarrow \Box_b {\Box_b^<}\Box_b \varphi$  is a theorem of in $\mathsf{P}_{\alg{B}}$. Then, by Interaction 1, we get that  
\begin{equation}\label{eq:axiom}
{\Box_b^<} \varphi \rightarrow \Box_b {\Box_b^<}\Box_b \varphi 
\end{equation} 
 is a theorem of $\mathsf{P}_{\alg{B}}$ as well.

For each element $b \in B^+$, we can see the restriction of the canonical model to $P^c_b$ as a set of $P^c_b$-clusters, namely maximal sets $C^b$ of elements from $W^c$ with respect to $P^c_b \cup (P^c_b)^{-1}$, i.e., such that for any $v,w \in C^b$, both $P^c_b(v,w)$ and $P^c_b(w,v)$. Any failure of condition C1 implying C2-a (i.e, elements in the canonical model for which both $P^{c<}_b(v,w)$ and $P^c_b(w,v)$) implies, by Inclusion axioms, that this happens inside some $P^c_b$-cluster. 
The following shows that this failure happens in fact inside $P^<_b$-clusters. The latter, analogously to $P^c_b$-clusters, are maximal sets of elements from $W^c$ with respect to $P^{c<}_b \cup (P^{c<}_b)^{-1}$.

\begin{lemma}(c.f. \cite[Lemma 1]{vBGiRo09}) \label{lemma:strictClusters}
	Let $C^b$ be a $P^c_b$-cluster in $\mod{M}^c$. If $P_b^{c<}(v,w)$ for some worlds $v,w \in C^b$, then $P_b^{c<}(s,t)$ as well for all $s,t \in C^b$.
\end{lemma}
\begin{proof}
	Take any $s \in C^b$, and any formula $\varphi$ such that $s(\Box^{<}_b\varphi) = 1$. Then, since the above formula \eqref{eq:axiom} is a theorem in $\mathsf{P}_\alg{B}$, we have $s(\Box_b{\Box_b}^{<}\Box_b\varphi) = 1$ as well.
	Now, since $s,v \in C^b$ we know $P^c_b(s,v)$, so $v({\Box_b}^{<}\Box_a\varphi) = 1$ by definition of $P_a$. By assumption, $P_b^{c<}(v,w)$, so $w(\Box_b\varphi)=1$, again by definition of $P_b^{c<}$. Finally, given that $P^c_b(w,t)$ (since both worlds belong to the cluster $C^b$), we get $t(\varphi) = 1$, proving $P_b^{c<}(s,t)$.
\end{proof}

In order to avoid these situations, i.e.\ loops of the form $[P^{c<}_b(w, u_1), P^{c<}_b(u_1, u_2), \ldots, P^{c<}_b (u_n, w)]$, the bulldozing construction creates $\mathds{Z}$ copies of each world in a $P^{c<}_b$-cluster, and then orders them strictly, mimicking the original behavior of the cluster but effectively removing any reflexivity over $P^{c<}_b$.

The construction of the bulldozed model is done in a similar way to \cite{vBGiRo09}. The only relevant difference is,  when ordering the new created worlds, to choose an ordering that takes into account the possible interactions of strict clusters for different level-cuts. This can be achieved thanks to the $\wedge$-transitivity of the $P^{c<}_b$'s. We include in Appendix B the technical details of the construction and the correctness of the bulldozed model.

All the previous considerations allow us to state the desired completeness of $\bf B$-valued preference logic $\mathsf{P}_\alg{B}$ with respect to the intended semantics.

\begin{theorem}[Completeness of $\mathsf{P}_\alg{B}$]\label{completeness}
For any set of formulas $\Gamma \cup \{\varphi\} 
 \subseteq {\bf PFm}$,
\[\Gamma \vdash_{\mathsf{P}_\alg{B}} \varphi \text{ if and only if }\Gamma \Vdash_{\class{P}_{\alg{B}}} \varphi .\]
\end{theorem}

\begin{proof}
Soundness was proven in Lemmas \ref{lem:sound12} and \ref{lem:sound3}. 
As for completeness, if $\Gamma \not \vdash_{\mathsf{P}_\alg{B}} \varphi$, 
we know (see details in Appendix B) we can transform the canonical model $\mod{M}^c = (W^c, \{P^c_b: b\in B\}, \{P^{c<}_b: b \in B^+\}, e^c)$ into a new model 
$\mod{N} = \langle N, \{S_b\}_{b \in B}, \{S^<_b\}_{b \in B^+}, f\rangle$ such that
\begin{itemize}
\item $\langle N, \{S_b\}_{b \in B}, f\rangle$ is a preference model, 
\item For each $b \in B^+$ and $v,w \in N$ it holds $S^<_b(v,w)$ if and only if $S_b(v,w)$ and not $S_b(w,v)$. 
\item There is $v \in B$ such that $f(v, \Gamma) \subseteq \{1\}$ and $f(v, \varphi) < 1$. 
\end{itemize}
This proves the theorem. 
\end{proof}

\section{Back to fuzzy modalities}\label{sec:axiomNoCuts}

In the previous section, we have provided a complete axiomatic system $\mathsf{P}_\alg{B}$ for the graded preference modalities $\Box_b$'s,  $\Box^<_b$'s (and the definable $\Diamond_b$'s and $\Diamond^<_b$'s). Before, in Section \ref{sec:defs} we have seen that the original fuzzy modalities $\Box$ and $\Box^<$ can be expressed from them. Thus, the system 
$\mathsf{P}_\alg{B}$ 
can be considered as an axiomatization of the modalities $\Box$, $\Box^<$ and $A$ (defining also  $\Diamond, \Diamond^<$ and $E$) in an extended language. 

In this section, we will explore a way of getting the same logic for the graded modalities $\Box, \Box^<, A$ modalities, without relying on the cut modalities. $\Box_a, \Box^<_b$. This can be achieved by extending the original language with only one additional operation (instead of $\vert \alg{B}\vert$ modal operations), which is particularly relevant in possible future works studying cases with infinite algebra  of evaluation $\alg{B}$ (because in such a way, the language would still be kept finite, in contrast with the language arising from the cut modalities).

In order to obtain an axiomatization of the modal logic without the addition of cut modalities, it is possible to generalize an approach introduced in \cite{BoEsGoRo09} 
that allows us to remove them as primitive operators in the language.

Indeed, if we enrich our language with the well-known Monteiro-Baaz $\Delta$ connective (see e.g. \cite{Ha98}), the graded modalities $\Box_b$ and $\Box^<_b$ turn to be expressible in terms of the original modal operators $\Box$ and  $\Box^<$. In fact, the most natural definition is based on the corresponding $\Diamond$-operations, themselves definable from their respective $\Box$-ones (Proposition \ref{lem:interdef}).

Recall that the Monteiro-Baaz $\Delta$ operation over a linearly ordered MTL-chain $\alg{B}$ is the operation defined as
\[\Delta(b) =\begin{cases} 1 &\hbox{ if }b = 1\\ 0 &\hbox{ otherwise } \end{cases}\]
for all $b \in B$. 
In the following, we denote by $\varphi \approx \const{b}$ the formula $\Delta(\varphi \leftrightarrow \const{b})$. 

\begin{lemma}
For any formula $\psi$, the following equalities hold:
\begin{align*}
\Box_b \varphi \equiv_{\mathsf{P}_\alg{B}}  & \bigwedge\limits_{a \in B} (\Delta (\const{b} \rightarrow \Diamond(\varphi \approx \const{a})) \rightarrow \const{a}) \\
\Box^<_b \varphi \equiv_{\mathsf{P}_\alg{B}} &  \bigwedge\limits_{a \in B} (\Delta (\const{b} \rightarrow \Diamond^<(\varphi \approx \const{a})) \rightarrow \const{a})
\end{align*}

\end{lemma}
\begin{proof}
Let $\mod{M}$ be a $\alg{B}$-preference model. We will do the details for the first case, the proof for the strict modalities is analogous using $P^<$.

	As in \cite{BoEsGoRo09} is easy to see that 
	$e(v, \Diamond(\varphi \approx \const{a})) = \bigvee\limits_{w: e(w,\varphi) = a} P(v,w)$.
	Then 	
	\[e(v, \Delta (\const{b} \rightarrow \Diamond(\varphi \approx \const{a}))) = \begin{cases} 
	1, & \text{if } b\leq \bigvee\limits_{w: e(w,\varphi) = a} P(v,w)\\ 
	0, & \text{ otherwise}  
	\end{cases} \]
	Let us denote $\delta_a(\varphi)\coloneqq \Delta (\const{b} \rightarrow \Diamond(\varphi \approx \const{a}))$, and $S = \{a  \in B\colon b\leq \bigvee\limits_{w: e(w,\varphi) = a} P(v, w)\}$. Then the previous equality implies
	\begin{align*}
 e(v, \delta_a(\varphi)) \rightarrow \const{a}) = \begin{cases} a, & \text{if } a \in S \\ 1, &  \text{ otherwise}  \end{cases} 
	\end{align*}	
It is a simple calculation to see that 
	$S = \{e(w,\varphi) \colon b \leq P(v,w)\}.$ 
	Then, we conclude the proof, since  
	\begin{align*}
	e(v,  \bigwedge\limits_{a \in B} \delta_a(\varphi) \rightarrow \const{a})) =  \bigwedge S = \bigwedge\limits_{b \leq P(v,w)} e(w, \varphi) = e(v, \Box_b \varphi)
	\end{align*}
\end{proof}

It is nor clear how to produce an axiomatization for the fragment with only $\Box,  \Box^<$ and $A$ (and the corresponding definable dual $\Diamond$-like operations) of the logic $\Vdash_{\class{P}_\alg{B}} $ plus $\Delta$. In this way, we will be able to avoid using the graded modalities $\Box_b$ and ${\Box_b^<}$. 
In order to do so, it is first easy to provide an axiomatic system for the whole  logic $\Vdash_{\class{P}_\alg{B}} $ plus $\Delta$ by adding to $\mathsf{mM}_\alg{B}$ an axiomatization for $\Delta$ on $\alg{B}$ (see eg. \cite{Ha98}, \cite{ViEsGo16})
and the interaction $\Box-\Delta$ axioms 
\begin{align*}
\Delta \Box_b \varphi \rightarrow \Box_b\Delta \varphi \text{ for }b \in B, \quad \text{ and }\quad 
\Delta \Box^<_b \varphi \rightarrow \Box^<_b\Delta \varphi \text{ for all }b \in B^+
\end{align*}
These latter axioms are only necessary to prove the meta-rule $$\Gamma \vdash \varphi \Rightarrow \Box_b \Gamma \vdash \Box _b\varphi, $$ and the corresponding one for ${\Box_b^<}$. Having that, the completeness proof coincides with the ones done for the logics without $\Delta$, simply defining the worlds of the canonical model as homomorphisms into the algebra $\alg{B}$  extended with $\Delta$.
From here, it is clear that we can use the interdefinability of $\Box_b, \Diamond_b$ from $\Diamond$ proven above, and obtain in that way an axiomatic system complete with respect to the intended semantics, over the language with only the original modal operators and the new $\Delta$.

Nevertheless, this axiomatization is still complete with respect to a conservative expansion of the intended graded preferences logic, since the language has been expanded with $\Delta$. Being conservative, any deduction in the restricted language holds in the logic if and only if it holds in the intended preference logic, but removing $\Delta$ would still be an interesting problem to face, even though it is not clear if it can be solved (namely, if the logic with $\Box, \Box^<, \mathsf A$ has a finite axiomatization).

\section{Modeling fuzzy preferences on propositions}\label{sec:relations}

The preference models introduced above are a very natural setting to formally address and reason over graded or fuzzy preferences over non-classical contexts. They are similar to the (classical) preference models studied by van Benthem et. al in \cite{vBGiRo09}, but offering a lattice of values (and so, a many-valued framework) where to evaluate  both the truth degrees of formulas and the accessibility (preference) relation. The latter can be naturally interpreted as a graded preference relation between possible worlds or states (assignments of truth-values to variables). The question is then how to lift a (fuzzy) preference relation $\leq$ on worlds to (fuzzy) preference relations among formulas. 

In the classical case, for instance in \cite{vBGiRo09,EsGoVi18} the following six extensions are considered, where $[\varphi]$ and $[\psi]$ denote the set of models of propositions $\varphi$ and $\psi$ respectively:  

\begin{itemize}
\item $\varphi \leq_{\exists\exists} \psi$ iff $\exists u \in [\varphi], \exists v \in [\psi]$ such that $P(u, v)$
\item $\varphi \leq_{\exists\forall} \psi$ iff $\exists u \in [\varphi]$, such that $\forall v \in [\psi]$, $P(u, v)$
\item  $\varphi \leq_{\forall\exists} \psi$ iff $\forall u \in [\varphi]$, $\exists v \in [\psi]$ such that $P(u, v)$
\item $\varphi \leq_{\forall\forall} \psi$ iff $\forall u \in [\varphi]$ and $\forall v \in [\psi]$, $P(u, v)$
\item $\varphi \leq_{\exists\forall2} \psi$ iff $\exists v \in [\psi]$, such that $\forall u \in [\varphi]$, $P(u, v)$  
\item $\varphi \leq_{\forall\exists2} \psi$ iff $\forall v \in [\psi]$,  $\exists u \in [\varphi]$ such that $P(u, v)$
\end{itemize}  
\vspace{0.2cm}
Analogous expressions could be obtained by replacing $P$ by its strict counterpart $P^<$. However, not all these extensions can be expressed in our framework, even if we restrict ourselves to classical propositions and classical preference relations. For instance, we can indeed express the orderings $\leq_{\exists\exists}$ and $\leq_{\forall\exists}$ (and their strict counterparts) as follows:  
\vspace{0.1cm}
\begin{itemize}
\item $\varphi \leq_{\exists\exists} \psi :=\ E(\varphi \land \Diamond \psi)$ \; \quad $\varphi <_{\exists\exists} \psi := \E(\varphi \land \Diamond^< \psi)$ 
\item  $\varphi \leq_{\forall\exists} \psi := \A(\varphi \to \Diamond \psi)$ \quad $\varphi <_{\forall\exists} \psi := \A(\varphi \to \Diamond^< \psi)$
\end{itemize} 
\vspace{0.2cm}
but some others would need to consider the inverse preorder $P^{-1}$ of $P$ in the models or to assume the preorder $P$ be total, and some other are not  just expressible (see \cite{vBGiRo09}). On the other hand, not all the extensions of the weak orderings above are also equally reasonable, for instance some of them are not even preorders. This is not the case of  $\leq_{\forall\exists}$ and $\leq_{\forall\exists2}$, that are indeed preorders.  

In \cite{EsGoVi18} the authors have generalized the above classical definitions by allowing preference relations $P$ to be graded or many-valued (with values in a scale $\bf B$), while keeping the propositions Boolean. Then, the extensions of the above orderings become graded as well, by replacing $\forall$'s and $\exists$'s by $\bigwedge$'s and $\bigvee$'s respectively, for instance: 

\begin{itemize}
\item  $[\varphi \leq_{\exists\exists} \psi] = \bigvee_{u \in [\varphi]} \bigvee_{ v \in [\psi]} P(u, v)$
\item  $[\varphi \leq_{\forall\exists} \psi] = \bigwedge_{u \in [\varphi]} \bigvee_{ v \in [\psi]} P(u, v)$
\end{itemize}  
It is worth pointing out that these expressions formally coincide with the way the modal formulas $\E(\varphi \land \Diamond \psi)$ and $\A(\varphi \to \Diamond \psi)$ respectively are evaluated in a $\bf B$-preference model when the propositions $\varphi$ and $\psi$ are Boolean.

\begin{example} 
Continuing Example \ref{examp4}, we can use the preference relation between alternatives or worlds to illustrate concepts holding in our model. 
For instance, for $\varphi, \psi$ crisp-valued formulas, $[\varphi \leq_{\exists \exists} \psi] \geq \alpha$ if and only
 if there is some alternative where $\psi$ holds that is preferred with degree at least $\alpha$ to some alternative where $\varphi$ holds. Similarly,
  $[\varphi \leq_{\forall \exists} \psi] \geq \alpha$ if and only for each alternative where $\varphi$ holds there is some alternative where $\psi$ holds that is preferred at least in degree $\alpha$.
  
In particular, in the model from Example \ref{examp4}, it is easy to compute the following values:
  \begin{itemize}
\item $[f \leq_{\forall \exists} m] = 0.5 = [m \leq_{\forall \exists} f]$, expressing that, in general, the agent does not have a clear preference of fish over meat nor vice-versa.
\item $[f \leq_{\exists \exists} m] = 0.7$, while $[m \leq_{\exists \exists} f] = 0.8$, which indicate there is some alternative with fish strongly preferred (0.8) to some alternative with meat, but there is also some alternative with meat quite preferred (0.7) to some alternative with fish.
\item $[b \wedge m \leq_{\forall \exists} b \wedge f ]= 0.8$, that can be understood as a contextual preference, that is, fixing $b$ as context (meaning only alternatives where the agent is in the beach are considered), 
while in general fish was not preferred over meat as we saw above, if the agent is on the beach, fish is strictly preferred.
\end{itemize}
\end{example}

In the full many-valued case of the logic $\mathsf{P}_\alg{B}$, where both propositions and preference relations are valued on a (same) scale $\bf B$, the formulas 
\begin{center}
$ E(\varphi \land \Diamond \psi)$, \quad $ E(\varphi \land \Diamond^< \psi)$\\
 $A (\varphi \to \Diamond \psi)$, \quad  $A (\varphi \to \Diamond^< \psi)$ \\
 \end{center}
make full sense as graded generalizations of the $\leq_{\exists\exists}, <_{\exists\exists}$ and $\leq_{\forall\exists}, <_{\forall\exists}$ preference orderings respectively. We will keep using the same notations $\varphi \leq_{\exists\exists}\psi$, $\varphi <_{\exists\exists}\psi$, $\varphi \leq_{\forall\exists}\psi$, and $\varphi <_{\forall\exists}\psi$ to refer to the $\mathsf{P}_\alg{B}$-formulas  $E (\varphi \land \Diamond \psi)$, $E (\varphi \land \Diamond^< \psi)$, $A (\varphi \to \Diamond \psi)$ and $A (\varphi \to \Diamond^< \psi)$ respectively. 

Note that, since the modalities $E$ and $A$ are universal, the values of these formulas in a preference model do not depend on the particular worls where they are evaluated.

In particular, it can be shown that these generalisations of $\leq_{\forall\exists}$ and $<_{\forall\exists}$ satisfy the properties in the next lemma. 
\begin{lemma}  The following properties hold:
\begin{itemize}
\item[(i)] $\leq_{\forall\exists}$ is a reflexive and $\odot$-transitive relation on formulas, i.e. we have the following validities:

$\Vdash_{\class{P}_\alg{B}} \varphi \leq_{\forall\exists}\varphi $, 

$\Vdash_{\class{P}_\alg{B}} (\varphi \leq_{\forall\exists}\psi) \to ( (\psi \leq_{\forall\exists}\chi )  \to  (\varphi \leq_{\forall\exists}\chi ))$.

\item[(ii)]  $<_{\forall\exists}$ is $\odot$-transitive: 

$\Vdash_{\class{P}_\alg{B}} (\varphi <_{\forall\exists}\psi) \to ( (\psi <_{\forall\exists}\chi )  \to  (\varphi <_{\forall\exists}\chi ))$

\end{itemize}
\end{lemma}

\begin{proof}
\begin{itemize}
\item[(i)] Reflexivity of $\leq_{\forall\exists}$: $\A(\varphi \to \Diamond \varphi)$ is valid in $\class{P}_\alg{B}$, since $\varphi \to \Diamond\varphi$ (i.e.\ axiom $(\mathsf{4}\Diamond)$) is valid in $\class{P}_\alg{B}$. 

$\odot$-Transitivity of $\leq_{\forall\exists}$: one can show that
\begin{equation} \label{bb}
\A(\varphi \to \Diamond \psi) \odot \A(\psi \to\Diamond \chi) \to \A(\varphi \to \Diamond \chi)
\end{equation} 
is also a valid formula in $\class{P}_\alg{B}$. Namely, this follows by first showing that the following formula expressing a form of monotonicity for $\Diamond$ holds true in $\class{P}_\alg{B}$:
\[ \A(\varphi \to \psi) \to \A(\Diamond \varphi \to \Diamond \psi) .\]
This in turn leads to the valid formula $\A(\psi \to\Diamond \chi) \to \A(\Diamond \psi \to \Diamond \Diamond \chi)$, but since $\Diamond\Diamond \chi \to \Diamond \chi$ holds true in $\class{P}_\alg{B}$ (Axiom 4), we get 

\[\A(\varphi \to \Diamond \psi) \odot \A(\psi \to\Diamond \chi) \to \A(\varphi \to \Diamond \psi) \odot \A(\Diamond \psi \to\Diamond \chi), \]

and by axiom $K$ for $\A$, it follows the validity of 
\begin{equation*} \label{dd}
 \A(\varphi \to \Diamond \psi) \odot \A(\Diamond \psi \to\Diamond \chi) \to  \A(\varphi \to \Diamond \chi), 
\end{equation*} 
that directly allows us to show the validity  of  \eqref{bb}.

\item[(ii)] The proof is completely analogous to the case of $\leq_{\forall\exists}$.
\end{itemize}
\end{proof}

\begin{example} \label{ex73}
Still continuing with the preference model used in Example \ref{examp4} and the computations therein, we can now ask for instance to which degrees the agent prefers a light meal to a heavy meal  and viceversa, always according to the preference order $\leq_{\forall \exists}$:\footnote{Note that we obviate the specification of the world when evaluating the two preference statements as they ae independent of the world.} 

\begin{itemize}

\item $e(l \leq_{\forall \exists} h) = \bigwedge_{v \in W} e(v, l) \to e(v, \Diamond h) =$

$\min(0.8 \to 0.3, 0.2 \to 0.7, 0.8 \to 0.4, 0.2 \to  0.7) = 0.8 \to 0.3 = 0.5$

\item $e(h \leq_{\forall \exists} l) = \bigwedge_{v \in W} e(v, h) \to e(v, \Diamond l) =$

$\min(0.3 \to 0.8, 0.7 \to 0.6, 0.3 \to 0.6, 0.7 \to  0.4) = 0.7 \to 0.4 = 0.7$
\end{itemize}
Thus, in general, the agent  prefers a bit more a light meal to a heavy meal (0.7) than the other way round (0.5). 
\end{example}

It is clear then that in the frame of the $\log{P}_{\alg{B}}$ logic one can suitably encode (weak and strict) preferences of a fuzzy proposition $\psi$ over another $\varphi$ by the formulas $\varphi \leq_{\forall\exists}\psi$ and $\varphi <_{\forall\exists}\psi$ respectively. These preferences between propositions actually enjoy the properties of a fuzzy $\odot$-preorder in the case of $ \leq_{\forall\exists}$ while $<_{\forall\exists}$  is only $\odot$-transitive.

Moreover, once could express contextual or conditional fuzzy preferences. For instance, regarding the above example, we could be interested in  
evaluate the preferences between a light and a heavy meal ($l$ and $h$) given the agent finds himself at a beach place ($b$).  Here $b$ is taken as the context that restricts the set of possible worlds, that is, we are led to evaluate the preferences between $b \land l$ and $b \land h$.  In general. we can consider contextual preferences of the form $$\delta: \varphi \leq_{\forall\exists} \psi$$
standing for an abbreviation of $(\delta \land \varphi)   \leq_{\forall\exists} (\delta \land \psi)$, where $\delta$ is a (fuzzy) non-modal formula. Analogously for $\leq_{\exists\exists}, <_{\exists\exists}, $ and $<_{\forall\exists}$.

\begin{example}
Continuing Example \ref{ex73}, we  ask ourselves how much the values of the preference expressions $l \leq_{\forall \exists} h$ and $h \leq_{\forall \exists} l$ change when the context is that the agent is at a beach zone. That is, let us compute the values of the contextual expressions $b: l \leq_{\forall \exists} h$ and $b: h \leq_{\forall \exists} l$. We first compute the values of the new modalities:

\begin{center}
$\begin{array}{l|c|c|c|c|}
 P              & \texttt bf    & \texttt bm & \texttt cf  & \texttt cm \\
  \hline \texttt bf   & 1    & 0.5 & 0.5 &  0.5 \\
  \hline \texttt bm & 0.8 & 1    & 0.6 & 0.8 \\
  \hline \texttt cf   & 0.8 & 0.5 & 1   &  0.7\\
  \hline \texttt cm & 0.6 & 0.5 & 0.5 & 1 \\
  \hline
 \end{array}  
 $
 $\quad \circ \quad$
 $\begin{array}{|c|}
   b\land h  \\
  \hline 0.3    \\
  \hline  0.7  \\
  \hline  0 \\
  \hline 0  \\
  \hline
 \end{array}  
 $
 $\quad = \quad$
 $\begin{array}{|c|}
  \Diamond (b \land h)  \\
  \hline 0.3    \\
  \hline  0.7  \\
  \hline  0.2 \\
  \hline 0.2  \\
  \hline
 \end{array}  
 $

 \end{center}
\begin{center}
$\begin{array}{l|c|c|c|c|}
 P              & \texttt bf    & \texttt bm & \texttt cf  & \texttt cm \\
  \hline \texttt bf   & 1    & 0.5 & 0.5 &  0.5 \\
  \hline \texttt bm & 0.8 & 1    & 0.6 & 0.8 \\
  \hline \texttt cf   & 0.8 & 0.5 & 1   &  0.7\\
  \hline \texttt cm & 0.6 & 0.5 & 0.5 & 1 \\
  \hline
 \end{array}  
 $
 $\quad \circ \quad$
 $\begin{array}{|c|}
   b\land l  \\
  \hline 0.8   \\
  \hline  0.2  \\
  \hline  0 \\
  \hline 0  \\
  \hline
 \end{array}  
 $
 $\quad = \quad$
 $\begin{array}{|c|}
  \Diamond (b \land l)  \\
  \hline 0.8    \\
  \hline  0.6  \\
  \hline  0.6 \\
  \hline 0.4  \\
  \hline
 \end{array}  
 $

 \end{center}
and hence, we finally have: 
\begin{itemize}

\item $e(b: l \leq_{\forall \exists} h) = \bigwedge_{v \in W} e(v, b\land l) \to e(v, \Diamond (b \land h)) =$

$\min(0.8 \to 0.3, 0.2 \to 0.7, 0 \to 0.2, 0 \to  0.2) = 0.8 \to 0.3 = 0.5$ \\

\item $e(b: h \leq_{\forall \exists} l) = \bigwedge_{v \in W} e(v, b\land h) \to e(v, \Diamond (b \land l)) =$

$\min(0.3 \to 0.8, 0.7 \to 0.6, 0 \to 0.6, 0 \to  0.4) = 0.7 \to 0.6 = 0.9$ \\

\end{itemize}
Therefore, one can observe that, in the context of being at the beach, the preference for a light meal to a heavy meal has increased (0.9) while the preference for a heavy meal to a light meal keeps being the same (0.5). 

\end{example}

In this last example we have considered the context described by a two-valued formula ($b$), but nothing would change if the context would have been described by a genuine fuzzy formula. Moreover, note that  in the logic $\log{P}_{\alg{B}}$ one could also express somewhat more involved preference statements of the form ``the more I prefer fish to meat, the more I prefer light to heavy meals'' by means of the implication

$$ (m \leq_{\forall \exists} f) \to  (h \leq_{\forall \exists} l) .$$
Indeed, if such an implication is assumed to be true, then it forces the truth-value of the formula $h \leq_{\forall \exists} l$ (i.e.\ the degree to which I prefer light to heavy meals) to be grater or equal to the truth-value of $m \leq_{\forall \exists} l$ (i.e.\ the degree to which I prefer fish to meat). 

To be more precise,  the following graded version of modus ponens is valid in $\log{P}_{\alg{B}}$: 
$$ \overline{r} \to (m \leq_{\forall \exists} f), \overline{s}\to ((m \leq_{\forall \exists} f) \to  (h \leq_{\forall \exists} l))  \vdash_{\log{P}_{\alg{B}}} \overline{s \odot t} \to (h \leq_{\forall \exists} l) $$
for any truth-values $s, t \in B$. Note that the textual description above corresponds to this inference pattern in the particular case when $t = 1$, and hence when $s \odot t = s$. In the general case, we still have that the greater are $s$ and $t$, the greater is $s \odot t$.

\section{Conclusions and Future work}
The aim of this work is to provide a formal framework generalizing the treatment of preferences in the style of eg. \cite{vBGiRo09} to a fuzzy context. We have first presented an axiomatic system encompassing reflexive and transitive modalities plus universal operators, that is shown to be the syntactical counterpart of many-valued Kripke models with (reflexive and transitive) graded (weak) preference relations between possible worlds or states. It is based on considering the cuts of the relations over the elements of the algebra of evaluation, solving in this way some problems arising from \cite{ViEsGo17a}. 
We further consider the extension of the previous logical system to cases when strict preferences (associated to the previous weak preferences) are taken into account. We propose an axiomatic system complete with respect to this intended semantic. 
We also show how to axiomatize the previous logics without relying in cutting the relations over the elements of the algebra, but instead expanding the language with only one new operation, the projection connective $\Delta$. This logical framework stands towards the use of modal many-valued logics in the representation and management of graded preferences, in the analogous fashion that (classical) modal logic has served in the analogous Boolean preference setting. 

In solving the previous questions, we close in a positive way an open problem from \cite{BoEsGoRo11} concerning the inter-definability of modal operators on the minimal modal logic over a finite residuated lattice.

It is still fairly unexplored the use of this framework to model graded preferences.  We have presented several examples to illustrate some of the possibilities the proposed logical setting offers, and partially developed the study of graded preferences between propositions in Section \ref{sec:relations}, but still posing many challenges and open questions. Further, we consider it could be interesting to observe the previous formal systems under the light of, instead modeling preference relations, serving as a framework of cost/pay-off related systems, relating the cost of certain executions in a given configuration (i.e., evaluation of some formula in some world of a model) and the cost of changing to a different configuration (i.e., the weight of the accessibility relation). Interestingly enough, the strict modalities also enjoy a natural counterpart, forcing a change of configuration at each moment of the execution. 

From a more theoretical point of view, the study of the previous systems over other classes of algebras of truth-values (e.g. including infinite algebras like those defined on the real unit interval $[0, 1]$ underlying {\L}ukasiewicz, Product or G\"odel fuzzy logics) 
seems also of great interest, both from a theoretical point of view and towards the modelization of situations needing of continuous sets of values. 

\subsection*{Acknowledgments}
The authors are thankful to an anonymous reviewer for his/her useful comments that have helped to improve the layout of the paper. Vidal has been supported by the grant no. CZ.02.2.69/0.0/0.0/17 050/0008361 of the Operational programme Research, Development and Education of the Ministry of Education, Youth and Sport of the Czech Republic, co-financed by the European Union. Esteva and Godo acknowledge partial support by the Spanish FEDER/ MINECO project RASO (TIN2015-71799-C2-1-P). 


\bibliographystyle{alpha}

\appendix

\section{Appendix: Minimal modal logics of a finite residuated lattice}

For the sake of being self-contained, in this appendix we recall from \cite{BoEsGoRo11} the main components of the minimal modal logic over a finite residuated lattice $\bf B$, and of the modal logic considering only models with crisp accessibility relation. The logics axiomatized by Bou et. al in the previous paper is the $\Box$-fragment, but as we proved in Proposition \ref{lem:interdef}, in the language with constants (which is the case in this work, and also in \cite{BoEsGoRo11}) $\Diamond$ can be defined from $\Box$. Thus, the logics $\mathsf{M}_{\alg{B}}$ and $\mathsf{CM}_{\alg{B}}$ detailed below also axiomatize  the logic with both modalities.

We recall from Section \ref{sec:modalDefs} the basic propositional setting. We assume $\alg{B} = (B, \land, \lor, \odot, \to, 0, 1)$ is a {\em finite} (bounded, integral, commutative) residuated lattice, and we also consider its canonical expansion $\Al[B^c]$ by adding a new constant $\overline{b}$ for
	every element $b \in B$ (canonical in the sense that the interpretation of $\overline{b}$ in $\Al[B^c]$ is $a$ itself.) The
	logic associated with $\Al[B^c]$ is denoted by $\Logic[B^c]$, and its logical consequence relation $ \models_{\Al[B^c]}$ is defined in the usual way and specified in Section \ref{sec:modalDefs}.

The language of the minimum modal logic over $\Al[B^c]$ is defined as usual  from a set of propositional variables $\cal V$, 0-ary truth constants $\{\overline{b} : b \in B\}$, binary propositional connectives $\{ \land, \lor, \odot, \to \}$ and unary modal operator $\Box$. We let $\alg{MFm}$ be the set of formulas build inductively in the usual way, namely, 
\begin{itemize}
\item Any variable and constant symbol is a formula, 
\item Given a formula $\varphi$, $\Box \varphi$ is a formula, and
\item Given two formulas $\varphi, \psi$, and any binary propositional connective $\star \in \{ \land, \lor, \odot, \to \}$, $(\varphi \star \psi)$ is a formula.
\end{itemize}
No other sequence of symbols is a formula. We will remove parentheses when they are redundant, and consider conjunction, disjunctions or products of indexed families of formulas (non-ambiguous in our logics because of the commutativity of these operators).

Some additional operations can be obtained as abbreviations of the original language, namely for $\varphi, \psi \in \alg{MFm}$, we let: 
\begin{itemize}
\item $\neg \varphi \coloneqq \varphi \rightarrow \const{0}$, 
\item $\varphi \leftrightarrow \psi \coloneqq (\varphi \rightarrow \psi) \odot (\psi \rightarrow \varphi)$, 
\item $\Diamond \varphi \coloneqq \bigwedge_{b \in B} (\Box (\varphi \rightarrow \const{a}) \rightarrow \const{a})$.
\end{itemize}

Kripke-style semantics for the modal logic is defined as follows.  
		An \emph{$\alg{B}$-Kripke model} is a triple $\mod{M} = \langle W, R, e \rangle$ where $W$ is a set of worlds,  $R\colon W \times W \to B$, is a $B$-valued accessibility relation between worlds, and $e\colon  W \times \mathcal{V} \to B$ is the evaluation of the model, which is uniquely extended to formulas as usual for the propositional connectives and for the modal operator by letting: 

			\[e(w, \Box \varphi) \coloneqq \bigwedge\limits_{w \in W}\{R(v,w) \to e(w, \varphi)\}\]

From Proposition \ref{lem:interdef}, we know that 
$e(v, \Diamond \varphi) = \bigvee\limits_{w \in W}\{R(v,w) \odot e(w, \varphi)\}$
	
		We let $\class{M}_\alg{B}$ denote the class of all $\alg{B}$-Kripke models, and the corresponding notion of (local) consequence relation will be denoted by $\Vdash_{\class{M}_{\alg{B}}}$.

The axiomatic system $\mathsf{M}_\alg{B}$  presented below is the logic denoted by $\Logic[\mathsf{Fr},B^c]$ in \cite[Def.~4.6]{BoEsGoRo11}, defined by:

		\vspace{-0.2cm}
		\begin{enumerate}
				\item an axiomatic basis for $\Logic[B^c]$ (see Appendix A from \cite{BoEsGoRo11})
				\item modal axioms for $\Box$: \\ \hspace*{1.1cm} $\Box \const{1}$, \quad \\ $(\mathsf{MD})$ \; $(\Box \varphi \land \Box \psi) \to \Box (\varphi \land \psi)$, \quad
				\\ $(\mathsf{Ax}_b)$ \; $\Box (\overline{b} \to \varphi) \leftrightarrow (\overline{b}
				\to \Box \varphi)$ 

			\item The rules of the basis for $\Logic[B^c]$ and the Monotonicity rule:  

	 \[(\mathsf{Mon}) \colon \text{ from }\varphi \to \psi \text{ derive }\Box \varphi \to \Box \psi.\]

		\end{enumerate}

 The corresponding notion of proof is denoted by $\vdash_{\mathsf{M}_\alg{B}}$.

	\begin{theorem}[(Th. 4.11, \cite{BoEsGoRo11}) Completeness of $\mathsf{M}_{\bf B}$]\label{Th:completeness}
		For any subset of formulas $\Gamma \cup \{ \varphi\}$, $\Gamma \vdash_{\mathsf{M}_{\alg{B}}} \varphi$ iff $\Gamma  \Vdash_{\class{M}_{\alg{B}}} \varphi $.
	\end{theorem}

In the case $\bf B$ is a finite MTL-chain, consider the subclass $\class{CM}_{\bf B} \subseteq \class{M}_{\alg{B}}$ consisting of models $\langle W, R, e \rangle$ with $R$ being a {\em crisp} accessibility relation (namely, $R \subseteq W\times W$). Then, the corresponding logic is given by the system $\mathsf{CM}_{\bf B}$ obtained by extending  $\mathsf{M}_{\bf B}$ with the following two additional axiom: 
\begin{itemize}
\item Axiom $C \colon \Box (\overline{k} \lor \varphi)  \to \overline{k} \lor \Box \varphi$, where $k$ is the co-atom (i.e., the immediate predecessor of $1$ in the algebra) of $\bf B$; 
\end{itemize}
This axiomatic is the one denoted by $\Logic[\mathsf{CFr},B^c]$ in [Def.~4.16]\cite{BoEsGoRo11}.

Further, it is easy to see it is equivalent to the one obtained by adding the $K$ axiom
\[K \colon \Box(\varphi \to \psi) \to (\Box \varphi \to \Box \psi),\]
 and changing the (Mon) rule to the Necessitation rule: 
 \[N_\Box \colon \text{ from }\varphi \text{ derive } \Box \varphi\]

The completeness proof for $\mathsf{CM}_{\bf B}$ is analogous to the one for the logic 
\begin{theorem}[(Th. 4.22, \cite{BoEsGoRo11}) Completeness of $\mathsf{CM}_{\bf B}$]\label{Th:completeness} 
Let $\bf B$ be a finite MTL-chain. Then, 
		for any subset of formulas $\Gamma \cup \{ \varphi\}$, $\Gamma \vdash_{\mathsf{CM}_{\alg{B}}} \varphi$ iff $\Gamma  \Vdash_{\class{CM}_{\alg{B}}} \varphi $.
	\end{theorem}

\section{Appendix: Construction and correctness of the Bulldozed Model}\label{bulldozed}

We will see how the canonical model $\mod{M}^c$ from Section \ref{sbs:completeness} can be deformed so no $P^<_b$ cycles are present.
The following lemma shows how several strict clusters (i.e., maximal sets with respect to $P^<_b \cap (P^<_b)^{-1}$) interacting have a very well behaved structure.
\begin{lemma}\label{lem:nested}
Let $a \leq b \in A$ and $C^a, C^b$ be respectively a $P^<_a$ and a $P^<_b$ cluster in $W$ such that $C^a \cap C^b \neq \emptyset$. Then $C^b \subseteq C^a$. 

Moreover, if $a < b$, then for any $a < c < b$ there is a $P^<_c$ cluster $C^c$ such that $C^b \subseteq C^c \subseteq C^a$. 
\end{lemma}
\begin{proof}
The first part follows by the $\wedge$-transitivity of $P^<$. The second is due to the fact that, for $b \leq c$, any $P^<_b$ cluster is inside a $P^<_c$ cluster.
\end{proof}

 Before defining the Bulldozed model, allow us to introduce some sets of worlds from the canonical model and built from them.
\begin{enumerate}
\item Consider for each $b \in B$, each $P^<_b$ cluster, and index them by $\{C^b_i\}_{i \in I_b}$ for suitable families of indexes $I_b$;
\item 
Use the following recursive procedure to obtain a strict order $<^b_i$ for each one of the previous clusters $C^b_i$: 
{\quotation{
\textbf{Ordering($C^x_j$):}
 \begin{itemize}
 \item If there is no $a > x$ and $C^a_s$ such that $C^a_s \subseteq C^x_j$, let $<_j^x$ be any arbitrary strict ordering of  $C^x_j$ (any order is compatible with $P^<_x$ in it, from Lemma \ref{lemma:strictClusters}).
 \item Otherwise, let $y$ be the immediate successor of $x$ in $A$, and let $C^y_{j_1}, \ldots C^y_{j_k}$ all the (disjoint) different $P^<_y$ clusters inside $C^x_j$. Call \textbf{Ordering($C^y_{j_n}$)} for each of them, obtaining strict orderings $<_{j_1}^y, \ldots <_{j_k}^y$ of each of the previous sub-clusters. Let then $<_j^x$ be  any ordering of the full $C^x_j$ compatible with each $<_{j_n}^y$. This ordering exists because the subclusters are all disjoint by definition, and fully contained in $C^x_j$ due to Lemma \ref{lem:nested}. Further, it is compatible with $P^<_x$ in $C^x_j$, since any strict order is so  from Lemma \ref{lemma:strictClusters}.
 \end{itemize}
 }}

\item For each cluster $C^b_i$ let $T^b_i \coloneqq \{\langle v,n\rangle \colon v \in C^b_i, n \in \mathds{Z}\}$; It is possible that $T^a_i \cap T_j^b  \neq \emptyset$, for $a \neq b$ (if there are nested clusters), but $T^b_i \cap T^b_j = \emptyset$ whenever $i \neq j$. 
\end{enumerate}

The bulldozed model is $\mod{T} = \langle T, \{S_b\}_{b \in B}, \{S^<_b\}_{b \in B^+}, f\rangle$ where: 
\begin{itemize}
\item Let $W^-$ be the worlds of the canonical model that do not belong to any cluster, and then let $T = W^- \cup \bigcup_{b \in B^+} \bigcup_{i \in I_b} T^b_i$;
\item $f(v,p) = e(v,p)$ for any $v \in W^-$, and $f(\langle v,n\rangle_b, p) = f(v,a)$ for any other world in $T$;
\item For each $b \in B^+$, we will  first define the strict accessibility relations $S^<_b$, by considering the different cases:

\begin{itemize}
\item If either $v$ or $w$  belong to $W \setminus \bigcup_{i \in I_b} C^b_i$ (i.e., outside any $P^<_b$-cluster), the the relation $P_b^<$ held condition (1 implies 2a): it is simple to see that indeed, either $P^<_b(v,w) = 0$ or $P^<_b(w,v) = 0$, since otherwise both elements would belong to some $P^<_b$ cluster. Thus, we simply let
\[
S^<_b(v,w) \text{ if and only if } 
\begin{cases} P^<_b(v,w) &  v,w \in W^-\\
P^<_b(v,u) &  v \in W^-, w = \langle u,n\rangle\\
P^<_b(u,w) &  w \in W^-, v = \langle u,n\rangle
\end{cases}
\] 

\item If $v \in C^b_i$ and $w \in C^b_j$ for $i \neq j$ (i.e., to different $P^<_b$ clusters), they are also well-behaved in the sense that either $P^<_b(v,w) = 0$ or $P^<_b(w,v) = 0$ (otherwise, they would belong to the same cluster). Then, again, for any $m,n \in \mathds{Z}$,  let \[S^<_b(\langle v,n \rangle, \langle w,m\rangle) \text{ if and only if } P^<_b(v,w).\]
\item If $v,w \in C^b_i$, then define
\[S^<_b(\langle v,n \rangle, \langle w,m\rangle) \text{ if and only if } \begin{cases} n < m, \text{ or } \\
n=m \text{ and } v <_i^b w
\end{cases}\]
($<_i^b$ being the strict order within the cluster from point 2. above)
\end{itemize}
\item We let $S_0 = T^2$, and we let $S_b$ to be the reflexive closure of $S^<_b$, i.e., 
\[S_b(y,z)   \text{ if and only if }S^<_b(y,z) \text{ or } y=z \text{ (in }T).\]

\end{itemize}

We need to check that the bulldozed model is indeed behaving as the canonical one, and also, that in it, the conditions relating $S_b$ and $S^<_b$ hold. The fact that the resulting fuzzy relation $S(v,w) = max_{b \in A} S_b(v,w)$ holds the conditions of an $\alg{B}$-preference model follow from the fact that $S_b$ is reflexive (by its own definition) and $\wedge$-transitive (because the $S^<_b$ are).

In order to prove the equivalence of both models, let us introduce a mapping $\beta \colon W \rightarrow \mathcal{P}(T)$ pairing each world from $\mod{M}^c$ with all the ones it generates in $\mod{T}$ in the obvious way, namely:
\[\beta(w) \coloneqq \begin{cases} \{w\} & \text{ if } w \in W^-\\ \{\langle w,n\rangle \colon n \in \mathds{Z}\} & \text{ otherwise. }\end{cases}\]

It will be useful the following simple result, slightly stronger than the basic (classical) bisimilarity of $\mod{M}^c$ and $\mod{T}$.
\begin{lemma} Let $v \in W$. Then 
\begin{enumerate}
\item For any $w \in W$ such that $P_b(v,w)$ (resp. $P^<_b(v,w)$) and any $u \in \beta(w)$ there is $z \in \beta(w)$ with $S_b(u,z)$ (resp. $S^<_b(u,z)$). 
\item For any $u \in \beta(v)$ and any $z\in T$ such that $S_b(u,z)$ (resp. $S^<_b(u,z)$) there is $w \in W$ such that $z \in \beta(w)$ and $P_b(v,w)$ (resp. $P^<_b(v,w)$. 
\end{enumerate}
\end{lemma}
\begin{proof}
\begin{enumerate}
\item If either one of $v$ or $w$ is outside all $P^<_b$ clusters, or if they belong to different ones, any $z \in \beta(w)$ serves, since the relations were preserved from the canonical model. Otherwise, we know $u = \langle v,n\rangle$ for some $n \in \mathds{Z}$. Then, for any $n < m$ it holds that 
$S^<_b(\langle v,n\rangle, \langle w,m \rangle)$ (and thus, also  $S_b(\langle v,n\rangle, \langle w,m \rangle)$).
\item By definition of $T$, $z \in \beta(w)$ for some $w \in W$. Now, if one of $v,w$ did not belong to any $P^<_b$ cluster, or if they belonged to different ones, we know again by definition that $S_b(u,z)$ if and only if $P_b(v,w)$ (and the same for what concerns $S^<_b$ and $P^<_b$), proving the claim. 

Otherwise it means that both $v,w$ belonged to the same $P^<_b$ cluster, so trivially $P^<_b(v,w)$ and $P_b(v,w)$.
\end{enumerate}
\end{proof}

\begin{corollary} For any $v \in W$ and any $\varphi \in {\bf PFm}$,
\[\{e(w,\varphi)\colon P_b(v,w)\} = \{f(z,\varphi) \colon S_b(u,z)\}, \text{ for any }u \in \beta(v)\]
\end{corollary}
\begin{proof}
It follows by induction on the complexity of $\varphi$, being the initial step due to the previous lemma and the definition of $f$ for propositional variables.
\end{proof}

From the previous, it trivially follows that the bulldozed model behaves like the original one, namely:

\begin{lemma}
For any formula $\varphi \in {\bf PFm}$ and any world $v \in W$, 
$e(v, \varphi) = f(u, \varphi)$ for any $u \in \beta(v)$.
\end{lemma}

We only need to check that indeed the new model is a preference model. Reflexivity and transitivity of $S_b$ follow by definition and by transitivity of $S^<_b$ (which holds by definition in the additional sets of worlds, and because $P^<_b$ was transitive itself).

\begin{lemma}
For any $v,w \in T$ and any $b \in B^+$ the following are equivalent:
\begin{enumerate}
\item $S^<_b(v,w)$ 
\item a) $S_b(v,w)$ and b) not $S_b(w,v)$. 
\end{enumerate}
\end{lemma}
\begin{proof}
Assume $v \in \beta(x), w \in \beta(y)$ (with possibly $x = y$). 

To show 1 implies 2, assume that $S^<_b(v,w)$. By definition of $S^<_b$, necessarily $v \neq w$. If either $x$ or $y$ did not belong to any $P^<_b$ cluster, or if they belong to different clusters, we know by definition that $P_b(x,y) = 1$, and also $P_b(y,x) = 0$ (otherwise, from Lemma \ref{lemma:strictClusters}, they would belong to the same  $P^<_b$ cluster). Since in this case we defined $S_b(v,w) = P_b(x,y)$ and $S_b(w,v) = P_b(y,x)$, this proves the implication. 
Suppose on the contrary that both $x,y$ belong to the same $P^<_b$ cluster. Then, in the way we defined $S^<_b$ for the elements of the unraveled cluster, there were no cycles, so $S^<_b(w,v) = 0$. Since $S_b$ in these worlds is the reflexive closure of $S^<_b$, we get that $S_b(v,w)$ and not $S_b(w,v)$.

To check that 2 implies 1, assume $S_b(v,w)$ and not $S_b(w,v)$. As before, if either $x$ or $y$ did not belong to any $P^<_b$ cluster, or if they belong to different clusters, we know that then $P_b(x,y)$ and not $P_b(y,x)$. 
Then, from Lemma \ref{lemma:twoImpliesone}, we know that $P^<_b(x,y)$, and thus $S^<_b(v,w)$ (again because $x,y$ belong to different $P^<_b$-cluster, so $S^<_b$ equals $P^<_b$). 

On the other hand, suppose both $x,y$ belong to the same $P^<_b$ cluster. In that case, $S_b$ is defined as the reflexive closure of $S^<_b$, so the assumptions imply that necessarily $S^<_b(v,w)$. 
\end{proof}

\end{document}